\documentclass[]{aastex7}
\usepackage{lipsum}
\usepackage{subcaption}
\usepackage{amsmath}

\begin{document}
			% yo creo que reconnection aporta confusión porque 
			% la gente puede pensar en la reconecxión magnética y eso
			% no es lo que se está haciendo acá
			
			\title{Lu and Hamilton model for solar flares over a rewiring complex network}

		\author[orcid=0000-0002-0211-3151,sname='Alejandro Zamorano']{Alejandro Zamorano}
		\affiliation{Departamento de F\'isica, Facultad de Ciencias, Universidad de Chile, Santiago, Chile}
		\email[show]{alejandro.zamorano@ug.uchile.cl}  
		
		\author[orcid=0000-0000-0000-0002,gname=Laura, sname='Laura Morales']{Laura Morales} 
		\affiliation{INFIP, UBA CONICET, Ciudad Aut\'onoma de Buenos Aires, Argentina}
		\affiliation{Ciclo Básico Común y Departamento de F\'isica, Universidad de Buenos Aires, Ciudad Aut\'onoma de Buenos Aires,
			Argentina}
		\email{lmorales@df.uba.ar}
		
		\author[orcid=0000-0002-4857-1051,gname=Denisse,sname=Denisse]{Denisse Past\'en}
		\affiliation{Departamento de F\'isica, Facultad de Ciencias, Universidad de Chile, Santiago, Chile}
		\email{
			denisse.pasten.g@gmail.com}
		
		\author[orcid=0000-0003-1746-4875,gname=Victor,sname=Victor]{V\'ictor Mu\~noz}
		\affiliation{Departamento de F\'isica, Facultad de Ciencias, Universidad de Chile, Santiago, Chile}
		\email{
			victor_munoz@uchile.cl}
		%% Use the \collaboration command to identify collaborations. This command
		%% takes an optional argument that is either a number or the word "all"
		%% which tells the compiler how many of the authors above the command to
		%% show. For example "\collaboration[all]{(DELVE Collaboration)}" wil include
		%% all the authors above this command.
		%%
		%% Mark off the abstract in the ``abstract'' environment. 
		\begin{abstract}
			We present a modified Lu \& Hamilton-type model where the neighborhood relations are replaced by topological connections, which can be dynamically altered. The model represents each grid node as a flux tube, as in the classic model, but with connections evolving to capture the complex effects of magnetic reconnection. Through this framework, we analyze how the dissipated energy distribution changes, particularly focusing on the power-law exponent $\alpha_E$, which decreases with respect to the original model due to rewiring effects. When the system is dominated by rewiring, it presents an exponential distribution exponent $\beta_E$, showing a faster decay of dissipated energy than in the original model.  This leads to microflare-dominated dynamics at short timescales, causing the system to lose the scale-free behavior observed in both the original model (Lu \&  Hamilton 1991) and in configurations where energy release is primarily driven by forcing rather than rewiring. 
			Our results reveal a clear transition from power-law to exponential regimes as the rewiring probability increases, fundamentally altering the energy distribution characteristics of the system. In contrast, when considering topological neighbors instead of local ones, the model's dynamics become intrinsically nonlocal. This leads to scaling exponents comparable to those reported in other nonlocal dynamical systems.
		\end{abstract}
		
		%% Keywords should appear after the \end{abstract} command. 
	%% The AAS Journals now uses Unified Astronomy Thesaurus (UAT) concepts:
	%% https://astrothesaurus.org
	%% You will be asked to selected these concepts during the submission process
	%% but this old "keyword" functionality is maintained in case authors want
	%% to include these concepts in their preprints.
	%%
	%% You can use the \uat command to link your UAT concepts back its source.
	
	\section{Introduction}\label{introduction}
	Solar flares are intermittent eruptive phenomena associated with rapid energy release in the solar corona. 
	The spatial coincidence of flares with magnetic structures at the solar surface supports the idea that flares are driven by nonpotential magnetic energy and
	triggered by an instability in the underlying
	magnetic configuration.  Moreover, their very 
	short onset time points to {\it fast magnetic reconnection} as the instability that converts magnetic energy into kinetic energy and radiation.
	%Flares have been observed consistently for several decades in  the extreme ultraviolet and soft and hard
	%X-ray domains of the electromagnetic spectrum.
	The energy range of solar flares has been systematically studied for almost the last $5$ solar cycles, and observations showed that their frequency distribution follows a well-defined power law, spanning eight orders of magnitude in flare energy:
	\begin{equation}
		f(E)=f_0\,E^{-\alpha}
	\end{equation}
	(see for example:~\cite{1985SoPh..100..465D}, 
	\cite{Aschwanden_2002} and~\cite{2022ApJ...934L...3A}
	with an excellent summary of observations in the last decades).
	
	Ultimately, solar flares can be interpreted as a self-similar phenomena that ({\it via} magnetic reconnection) heat the solar corona.  E. N. Parker conjectured these two features could be integrated in a simplified but plausible scenario in which a coronal loop, composed of a set of field lines nicely parallel to one another contained within a bent cylinder, subject to stochastic displacement of photospheric footpoints by convective motions will, at some point, lead to a situation where field lines are highly entangled and distorted~(\cite{1988ApJ...330..474P} \&~\cite{1983ApJ...264..642P}).
	Moreover, the high electrical conductivity of the coronal
	plasma facilitates the formation of tangential 
	discontinuities in these distorted regions: {\it current sheets}. 
	As the currents build up, magnetic reconnection sets in, field lines reconfigure, and dissipation occurs within the aforementioned current sheets. This is a collective behaviour where many current sheets continuously reconnect throughout the corona. ~\cite{1988ApJ...330..474P} estimated the energy release of one of those elementary reconnection events at $\sim 10^{24}$ erg and assumed that the addition of all the individual events could amount to the whole heating of the solar corona.
	
	Concurrently, Bak and collaborators developed the new self-organized criticality paradigm (hereafter SOC; see~\cite{1987PhRvL..59..381B},~\cite{1989PhRvA..39.6524K},~\cite{Jensen1998-JENSCE}, and references therein) to study the formation of avalanches in a sand pile. They found that the avalanche size spectrum has the form $f(n)=n^{-\alpha}$ (where $n$ is the number of sand grains involved in an avalanche).  In general, power laws may be indicative of scale-invariant dynamics, SOC dynamics being one of the ways to generate scale invariance (see, e.g.,~\cite{1999PhRvE..59.6175M}, ~\cite{2004cpns.book.....S}), and 
	can be observed in a wide range of physical systems exhibiting episodic activity: earthquakes and seismic noise emission, landslides, snow avalanches, forest fires, and, of course, solar flares~\cite{paczuski1996universality,sornette1989self,hergarten1998self, malamud1999self,clar1996forest,lu1993solar} (for a recent discussion on the ``flicker noise" for the case of coronal loops observations, see~\cite{2014ApJ...795...48C}).
	
	\cite{1991ApJ...380L..89L}, LH91 from now on, conceived a way to turn Parker's coronal heating model into an avalanche model for flares of all sizes. 
	They conjectured that reconnection at one tangential
	discontinuity can alter physical conditions in the
	vicinity of the reconnection site so that the
	corresponding neighboring tangential discontinuities can be pushed beyond the instability threshold; further reconnection at some of these sites can then trigger more reconnection events at other sites further away from the original reconnection site, and so on along and across the loop, until stability has been restored everywhere. The energy released by the ensemble of tangential discontinuities having undergone reconnection is an avalanche of events that they interpreted as a flare.
	
	Since the seminal work of LH91, the study of
	avalanches as solar flares has led to multiple
	models in $2$D and $3$D, with various boundary conditions (for a deep review see~\cite{2001SoPh..203..321C} and
	chapter $12$ of~\cite{2013socs.book.....A}).
	Most of them managed to provide statistical
	features for flares similar to those reported by observations~\citep{2000ApJ...535.1027A}. Nevertheless, none of those models managed to represent, explicitly, the magnetic reconnection process.  Other 
	works~\cite{2008ApJ...682..654M,hughes2003solar} have provided avalanche models involving reconnection of magnetic lines. In this article, we present a
	hybrid model for solar flares that integrates the canonical approach of LH91, and rewiring of a complex network, as a representation of reconnection events.  Links between sites define the possible routes for energy dissipation, so that the usual dynamics of LH91 occurs not on a regular grid defined by spatial neighbors, but over a complex network where neighborhood is defined in a topological sense.  {A key distinction of our approach, compared to the model of nonlocal communication by \cite{mackinnon1997nonlocal}, lies in the nature of nonlocality itself. Their model employed a static rule for nonlocal connections (e.g., influencing a predefined number of random sites), which implied a change in the effective magnetic topology. In contrast, our model explicitly incorporates dynamic topological evolution as the fundamental driver of nonlocal interactions}. Thus, the issue of energy distribution during solar flares becomes similar to the issue of transport over a network with varying topology, a problem which has been previously explored for other systems, where the relevant quantity being transported may be not only energy, but money, people, cars, etc.~\cite{munoz2022wealth, guillier2017optimization, lin2013complex, danila2006optimal}. In general, complex networks have been increasingly used to study various space and astrophysical systems, including solar flare dynamics, sunspots evolution, and solar wind dynamics~\cite{flandez202522, munoz2022complex, zurita2023characterizing, daei2017complex, tajik2024complex, gheibi2017solar}, as they provide a powerful framework for characterizing the nonlinear behavior, by means of network topology. In this work, instead of describing the system, the network in an intermediary over determining the energy redistribution pathways of energy flow. This is described in detail in Section~\ref{model}, where we also study how the system reaches a statistical stationary state.
	Then, in Section~\ref{results} we characterize the statistical properties of the avalanches produced by the proposed model. Finally, we summarize and discuss our findings in Sec.~\ref{summary}. 
	\section{The Lu-Hamilton model}
	\label{model}
\begin{figure}
		\centering
		\includegraphics[width=0.5\linewidth]{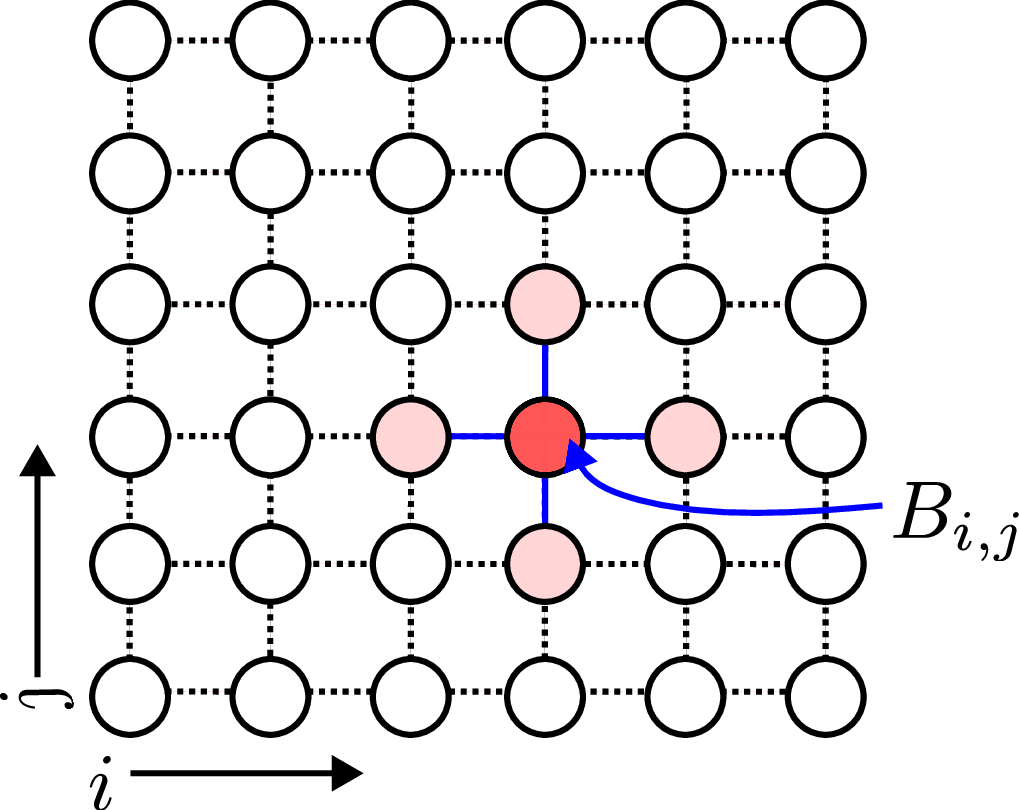}
		\caption{2D Cartesian model representing the original LH91 model. Each grid point is associated with a field $B_{i,j}$, and with four neighbors in vertical and horizontal directions as shown.}
		\label{fig:lhrepresentation}
	\end{figure}

	The LH model proposed by~\cite{1991ApJ...380L..89L} is a cellular automaton model of solar flares that represents coronal loops as a grid of interacting magnetic flux tubes. Each node $(i,j)$ in the grid is characterized by its magnetic field vector $\mathbf{B}_{i,j}$ (see Fig.~\ref{fig:lhrepresentation}). The system evolves through stochastic driving, where a randomly selected node receives a magnetic field increment $\delta\mathbf{B}$ with components uniformly distributed in $[-0.2, 0.8]$. This asymmetric driving causes gradual energy accumulation until the system reaches a self-organized critical (SOC) state.
	The instability condition is triggered when the magnetic field difference $d\mathbf{B}$ between neighboring flux tubes exceeds a critical threshold $B_c$, {\it i.e.\/}
	\begin{equation}
		\label{threshold}
		\|d\mathbf{B}\| > B_c \ . 
	\end{equation} This leads to localized energy redistribution through magnetic reconnection, modeled in the original 3D formulation as:
	
	\begin{equation}
		\mathbf{B}_i \to \mathbf{B}_i - \frac{6}{7}\text{d} \mathbf{B}_i\;,\quad \mathbf{B}_{nn}\to \mathbf{B}_{nn}+\frac{1}{7}\text{d}\mathbf{B}_i\;,
	\end{equation}
	where $\mathbf{B}_{nn}$ denotes the nearest-neighbor fields. 
	The generalization of this model to more dimensions is given by
	\begin{equation}
		\mathbf{B}_k \to \mathbf{B}_k - \frac{2D}{2D+1}Z_c\;,\quad \mathbf{B}_{nn}\to \mathbf{B}_{nn}+\frac{1}{2D+1}Z_c\;,
	\end{equation}
	where $D$ is the dimensionality ($D=2$ in our case), $Z_c$ is the critical magnetic field threshold ($Z_c=1$) When the field difference $\text{d}\mathbf{B}$ between neighboring flux tubes exceeds $Z_c$ (i.e., $|\text{d}\mathbf{B}| > Z_c$), the system undergoes localized energy redistribution. The dissipated energy $E$ of the system is calculated adding up all the energy dissipations until the energy change in the system returns to zero, i.e., the avalanche ends.  The minimum energy released is defined by $e_0=\frac{2D}{2D+1}Z_c^2$, and the energy of the whole system is the sum magnetic energy $|\mathbf{B}|^2$ of every node.

	\section{Complex network approach} 
	
\begin{figure}[ht!]
		\centering
		\hfill
		\begin{subfigure}[b]{0.25\textwidth}
			\centering
			\includegraphics[width=\linewidth]{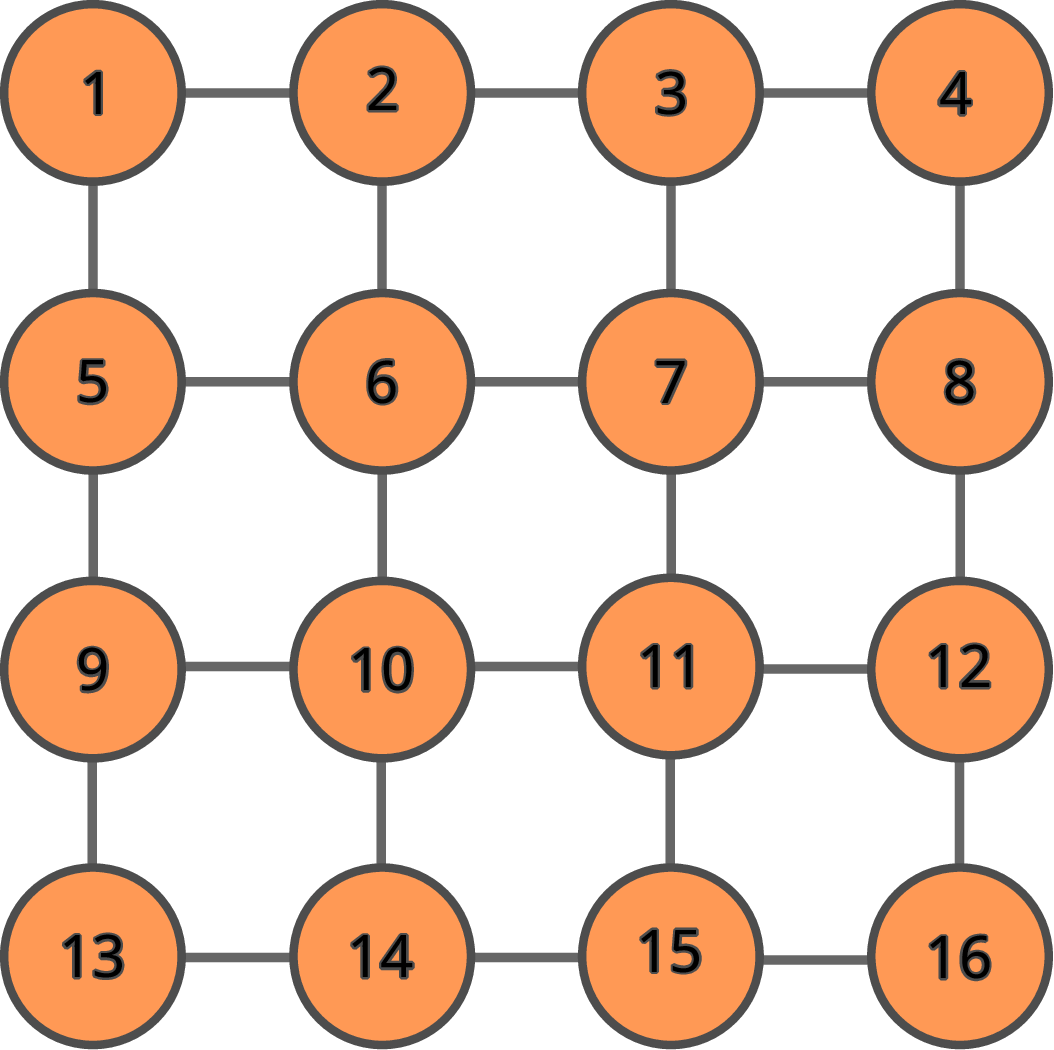}
			\caption{}
			\label{fig:estadocero}
		\end{subfigure}
		\hfill
		\begin{subfigure}[b]{0.25\textwidth}
			\centering
			\includegraphics[width=\linewidth]{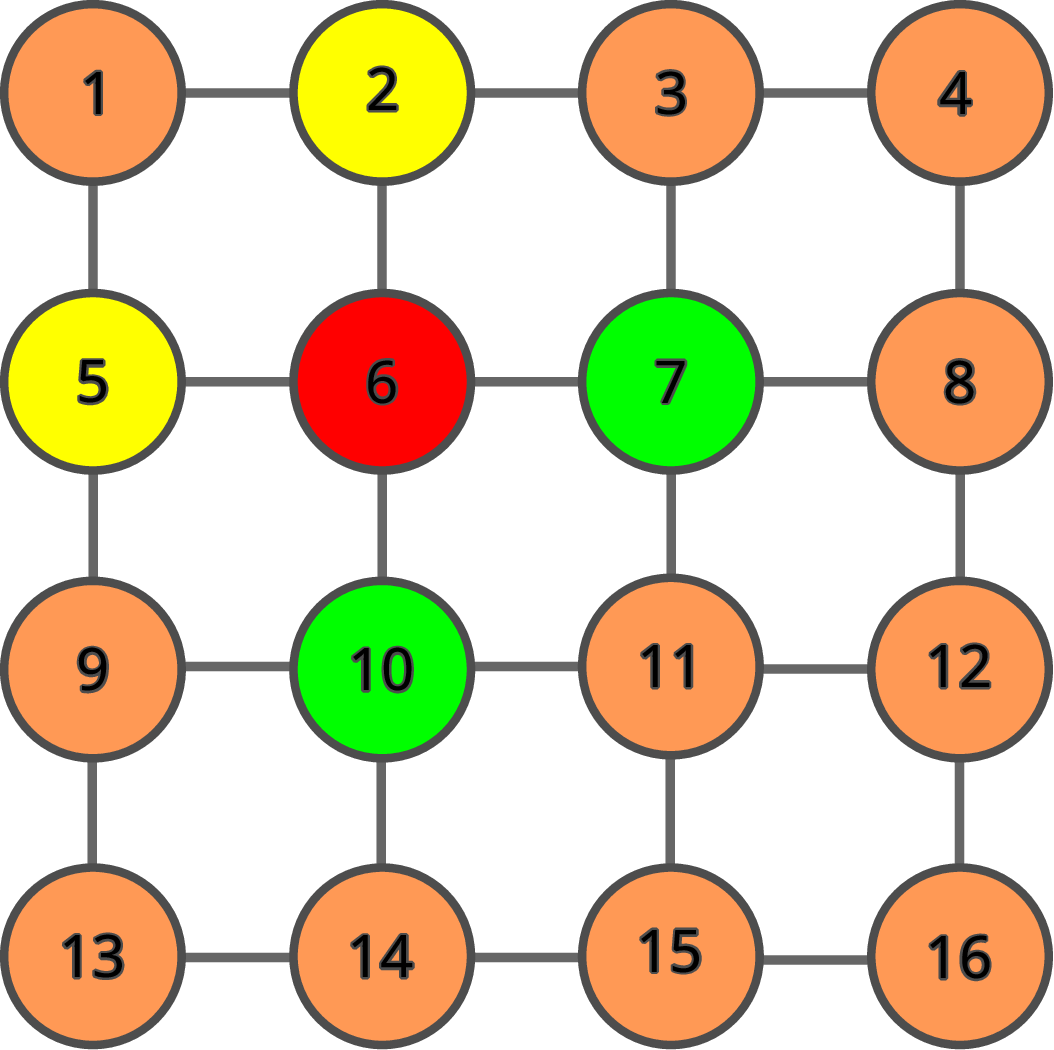}
			\caption{}
			\label{fig:estadouno}
		\end{subfigure}
		\hfill
		\begin{subfigure}[b]{0.25\textwidth}
			\centering
			\includegraphics[width=\linewidth]{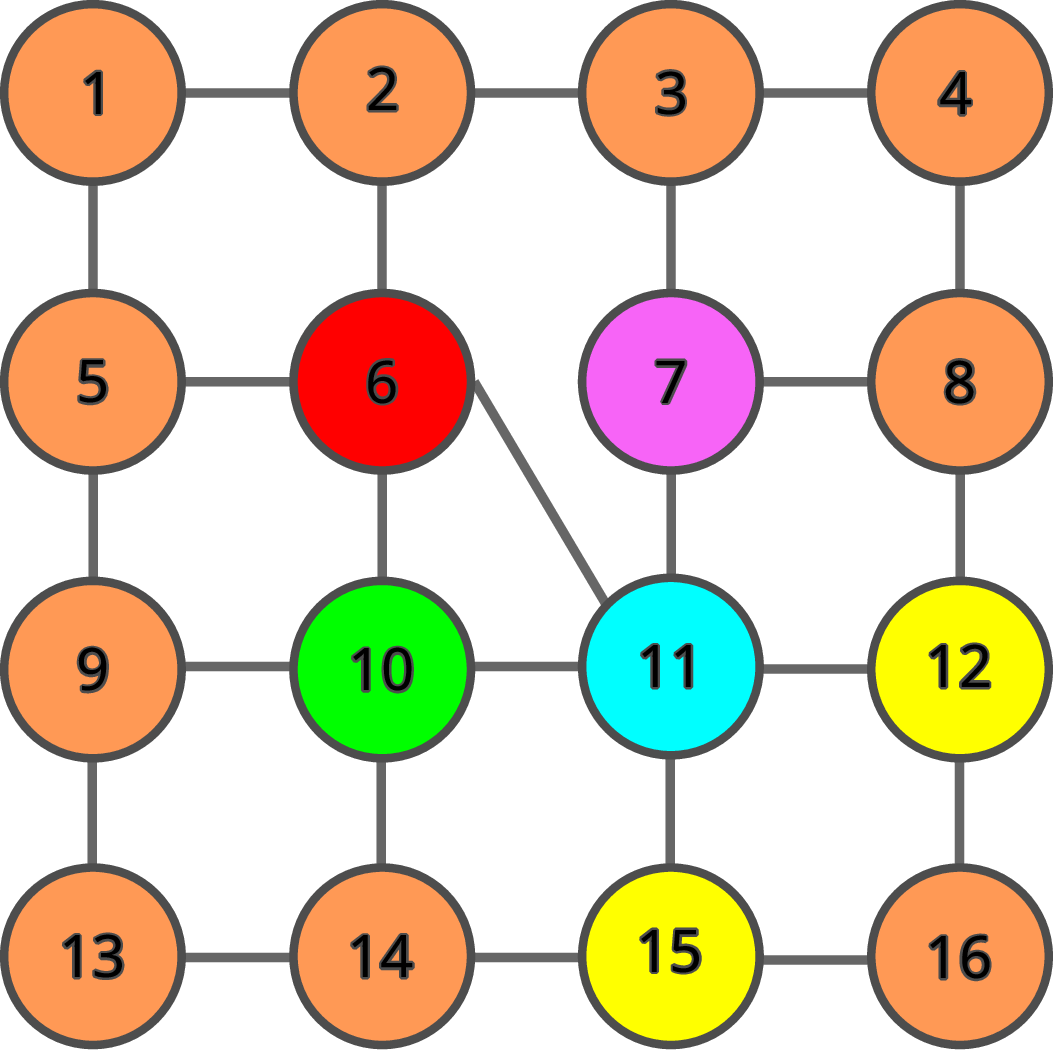}
			\caption{}
			\label{fig:estadodos}
		\end{subfigure}
		
		\begin{subfigure}[b]{0.25\textwidth}
			\centering
			\includegraphics[width=\linewidth]{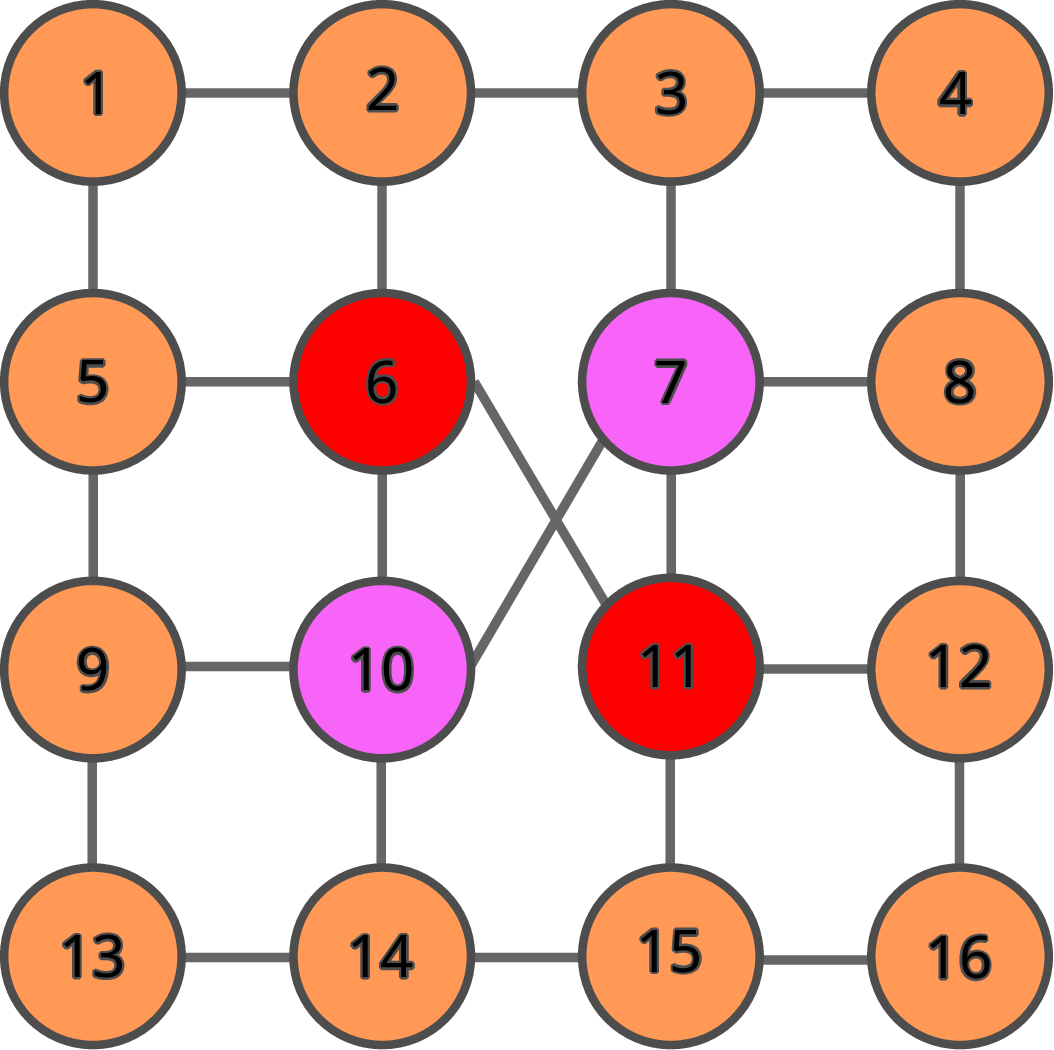}
			\caption{}
			\label{fig:estadotres}
		\end{subfigure}
		\hspace*{2cm}
		\begin{subfigure}[b]{0.25\textwidth}
			\centering
			\includegraphics[width=\linewidth]{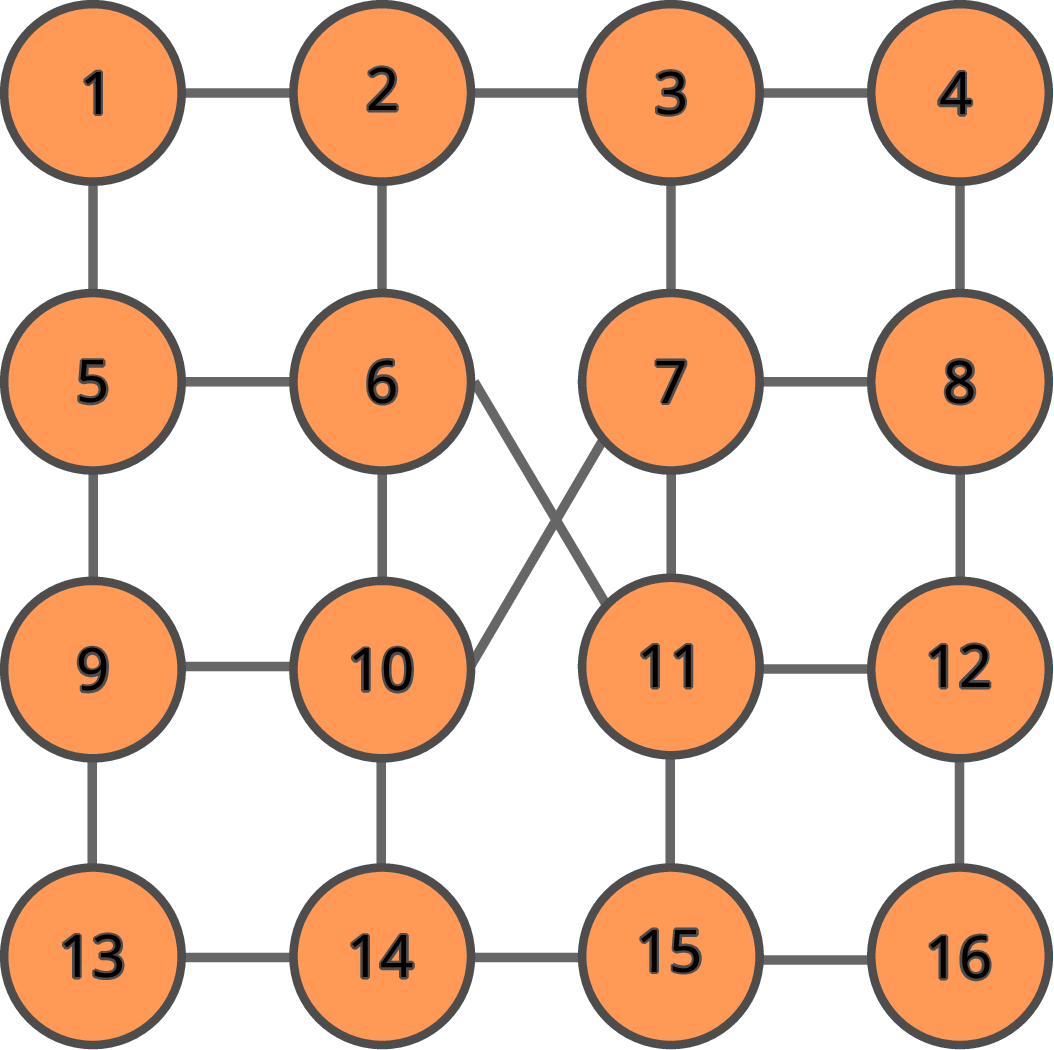}
			\caption{}
			\label{fig:estadocuatro}
		\end{subfigure}
		
		\caption{Diagram of the system where  (a) is the original lattice before the rewiring processes. In (b) a random node is selected, and its neighbors are identified; the selected node is red, yellow nodes are neighbors which cannot be selected because they are at the boundary, green nodes are neighbors which can be selected. In (c), one of the green nodes has been selected to move its connection to another, randomly selected node; the magenta node is the node which has just lost a connection, and the cyan node is the node who has gained that connection. In (d), the rewiring has been carried out, conserving the number of neighbors, highlighting the modified nodes, and (e) shows the final, rewired state.}
		\label{rewiring}
	\end{figure}

	Since flare events are associated with reconnection events, they are related to a change of the topological structure of the magnetic field.  In order to follow the dynamics of the Lu-Hamilton model on a varying topology, we will consider a complex network approach, where the regular grid of the LH91 model is regarded as a complex network. Then, each node represents a magnetic field line, an vertices of the network connect each node to its neighbors, thus defining the paths for energy redistribution when a flare is triggered.  
	For our $2D$ implementation, we use generalized redistribution rules. The $N-$dimensional topological rules are given by:
	\begin{equation}
		\mathbf{B}_k \to \mathbf{B}_k - \frac{2D}{2D+1}Z_c\;,\quad \mathbf{B}_{tn}\to \mathbf{B}_{tn}+\frac{1}{2D+1}Z_c\;,
	\end{equation}
	$\mathbf{B}_{tn}$ is the field at topological neighbors, which are the nodes connected to node $k$, and not necessarily the nearest neighbors in the spatial sense.
	Initially, the network is a direct representation of the regular grid, where vertices connect nearest neighbors. Thus, in this case, topological neighbors are identical to geometrical neighbors (see Fig.~\ref{fig:estadocero}). This cellular automaton has the same dynamics as in the LH91 model. 
	
	We now represent the changes in topology by rewiring the complex network, that is, changing the nodes that a given vertex connects to. This means that topological neighbors are not necessarily nearest neighbors in the geometrical sense, which in turn leads to a nonlocal dissipation process (although local in the topological sense). Some restrictions must be considered during the rewiring process, to maintain the physical meaning of the model. First, rewiring will conserve the number of neighbors of each node (node degree), in order to avoid the possibility that some node or set of nodes becomes isolated from the rest of the network, so that energy can always be redistributed along the whole original system. Second, rewiring will be such that loops are avoided. Thus, if energy flows from a certain node $A$ to another node $B$, for instance, it cannot flow back from $B$ to $A$. This forces energy to flow to the outer limits of the network, where it is eventually dissipated as in the LH91 model, avoiding possibly divergent energy accumulation inside the system. 
	
	%	a	
	Considering the restrictions outlined above, the rewiring scheme involves a four-step process, outlined below and represented in Fig.~\ref{rewiring}. Initially, the network is equivalent to a regular grid, as shown in Fig.~\ref{rewiring}(a). Step 1: A certain node is selected for rewiring [the red node in Fig.~\ref{rewiring}(b)]. Step 2: One of its connections is selected for deletion. This connection must be to a neighbor in the bulk [green nodes in Fig.~\ref{rewiring}(b)], not in the edges [yellow nodes], so that energy dissipation at the boundaries is always guaranteed. Step 3: The deleted connection is replaced with a new connection to a randomly selected node [cyan node in Fig.~\ref{rewiring}(c)]. Step 4: After the previous step, one node has lost a connection (node 7, in magenta), and one has gained a connection (node 11). In order to conserve the number of neighbors, one of the extra connections of node 11 is randomly selected to replace the missing connection to node 7. As before, the connection to be deleted can only involve a neighbor in the bulk [in this case, the only possibility is node 10, green, in Fig.~\ref{rewiring}(c)], not in the edges [yellow nodes in Fig.~\ref{rewiring}(c)]. The final state is shown in Fig.~\ref{rewiring}(d). It can be observed that this process ensures that all nodes have the same degree as in the initial connection, but connections are not necessarily to geometrical neighbors.

	We are interested in quantifying the effect of rewiring in the network on the statistics of avalanches in the LH model. We notice that rewiring has the effect of changing the neighbors, and thus there could be nodes which would be stable in the regular grid, but would be unstable in the rewired grid, because \eqref{threshold} holds for their topological neighbors. This, in turn, means that avalanches could be triggered by rewiring, not only by the external driver. 
	
	In order to study this systematically, we drive the system as usual, increasing the value of $\mathbf B$ at a node, but, with a certain probability, we replace the driver with a rewiring event. In general, at a given time step, the probability of driving is $p_d$, and of rewiring is $p_r=1-p_d$. $p_d=1$ corresponds to the usual LH91 model, and $p_d=0$ is a system where only rewiring events occur. 
	
\begin{figure}[ht!]
		\centering
		
		\begin{subfigure}[b]{0.5\linewidth}
			\centering
			\includegraphics[width=\linewidth]{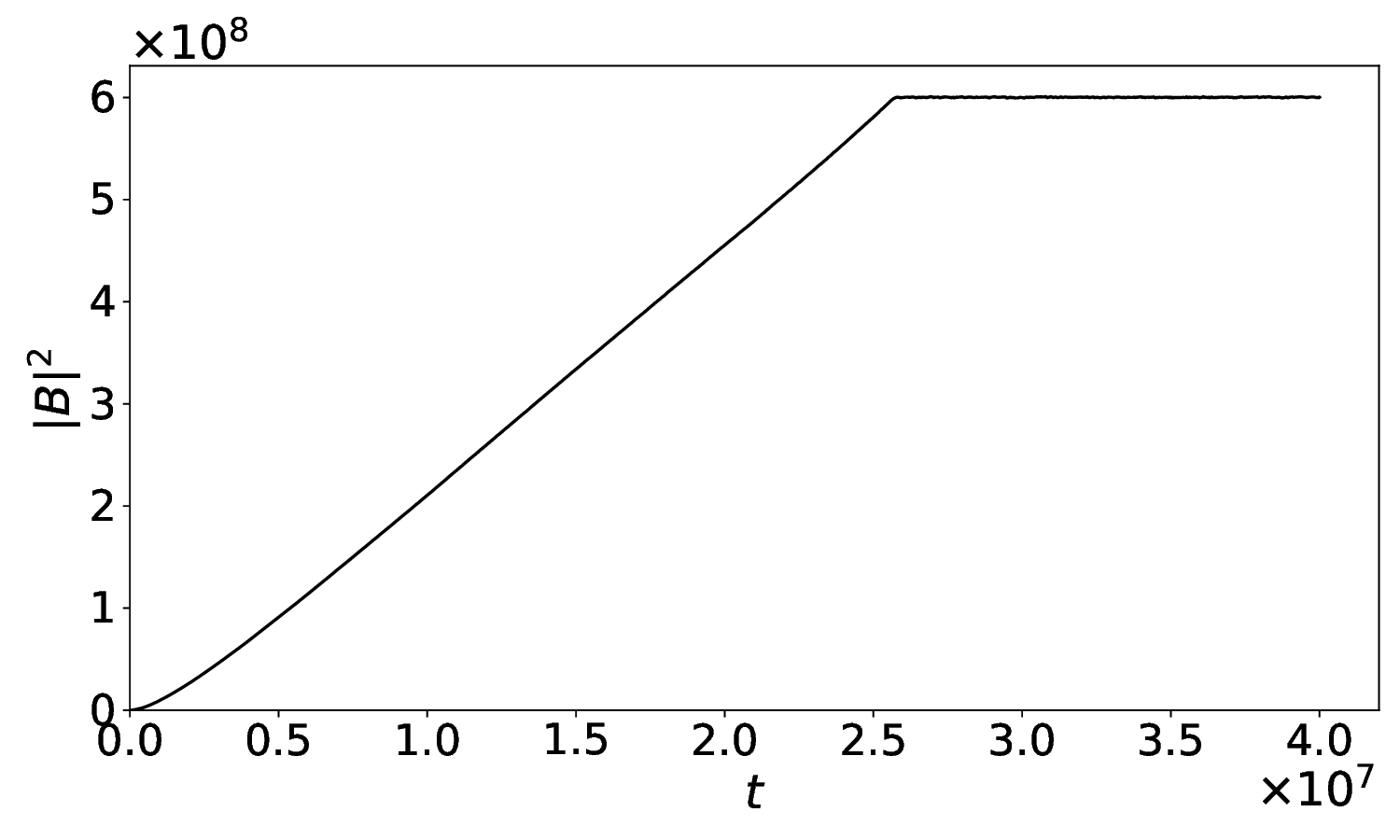}
			\caption{}
			\label{fig:lhenergy}
		\end{subfigure}
		
		\vspace{1em} % Espacio vertical entre las dos subfiguras
		
		\begin{subfigure}[b]{0.5\linewidth}
			\centering
			\includegraphics[width=\linewidth]{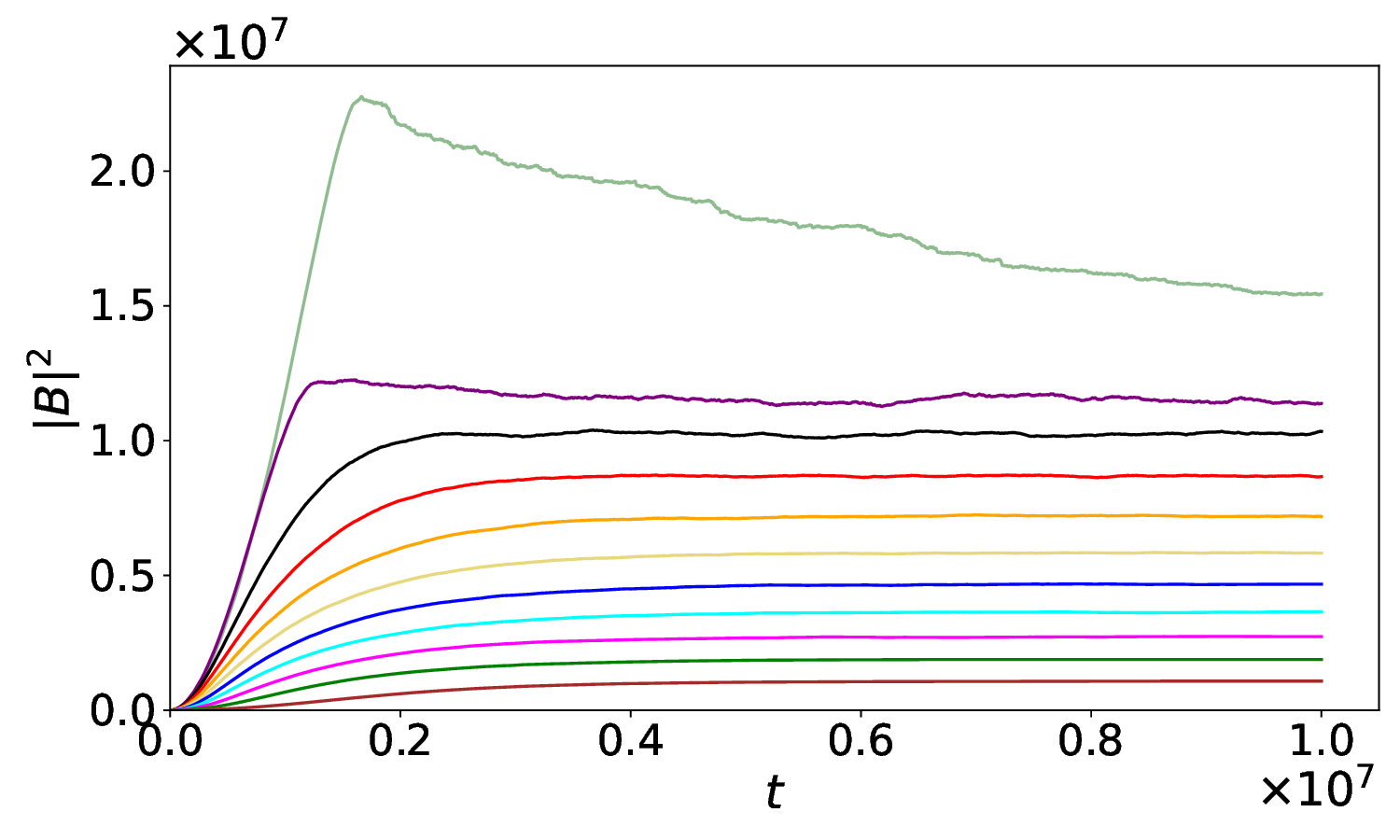}
			\caption{}
			\label{fig:all_energy}
		\end{subfigure}
		
		\caption{(a) LH total energy at time $t$. (b) Total energy for the whole system at time $t$, where the colors represent the different relations in percentage between driving-rewiring as a mechanism for the avalanche, i.e.  light green: 99.9-0.1, {purple: 99-1}, black: 90-10, red: 80-20, orange: 70-30, yellow: 60-40, blue: 50-50, cyan: 40-60, magenta: 30-70, green: 20-80, brown: 10-90. Using $10^7$ iterations and with $N=64$, $Z_c=1$.}
		\label{fig:combined_energy}
	\end{figure}

	We now consider a system with $N=64$ sites, so that the initial state corresponds to a square lattice of size 8, we start with a random initial condition (random field values at each site), and run the system for $10^7$ iterations, for a given driving-rewiring event ratio. Figure~\ref{fig:all_energy} shows the time series
	of the total lattice energy resulting from these simulations. Figure~\ref{fig:lhenergy} shows the result for the LH91 model ($p_d=1$), and Fig.~\ref{fig:all_energy} for several driving-rewiring ratios, from 90-10 ($p_d=0.9$), to 10-90 ($p_d=0.1$). We also include the cases 99-1 ($p_d=0.99$) and 99.9-0.1 ($p_d=0.999$), to discuss the transition, which occurs as soon as rewiring is introduced.

	It is apparent from the comparison of both
	plots that our model reaches a stationary state 
	$10$ times faster than LH91. 
	Moreover, the total energy that the system needs to accumulate in order to start generating avalanches of all sizes is one order of magnitude smaller for the rewired system than for LH91. {Fig.~\ref{fig:all_energy} shows that this occurs early when rewiring is introduced, as the decrease in total energy and in the time needed to reach a stationary state, is observed even for rewiring probability $p=0.01$.}
	This is clearly indicative that rewiring strongly hinders the building up of energy, since redistribution is facilitated by the paths opened by the rewiring scheme.

	Using the model described, we can study the statistical features of the avalanche process when rewiring is introduced in the Lu-Hamilton model. This is carried out in Sec.~\ref{results}.
	
	\section{Rewiring and Statistical Properties}
	\label{results}
	
	We first notice that, as in the usual LH91 model, avalanches occur as the system is driven.  However, since the avalanche condition depends on the state of neighboring sites, and when rewiring occurs neighbors change, we can expect avalanches to be triggered not only when energy is added to the system, but also due to the rewiring process. In the latter case, there is no net increment of the system's energy. 
	
	%Since the rewiring process involves four nodes, it can simultaneously induce two avalanches. It is important to %note that the system only gains energy through driving (via increments of $\delta B$), while rewiring %exclusively decreases the system's energy. Notably, neither process necessarily triggers an avalanche: energy %%increments $\delta B$ may occur without producing avalanches, and similarly, rewiring events can take place %without avalanche initiation. An avalanche is only considered to have occurred when it is actually observed in %the system.
	
\begin{figure}[ht!]
		\centering
		\includegraphics[width=0.7\linewidth]{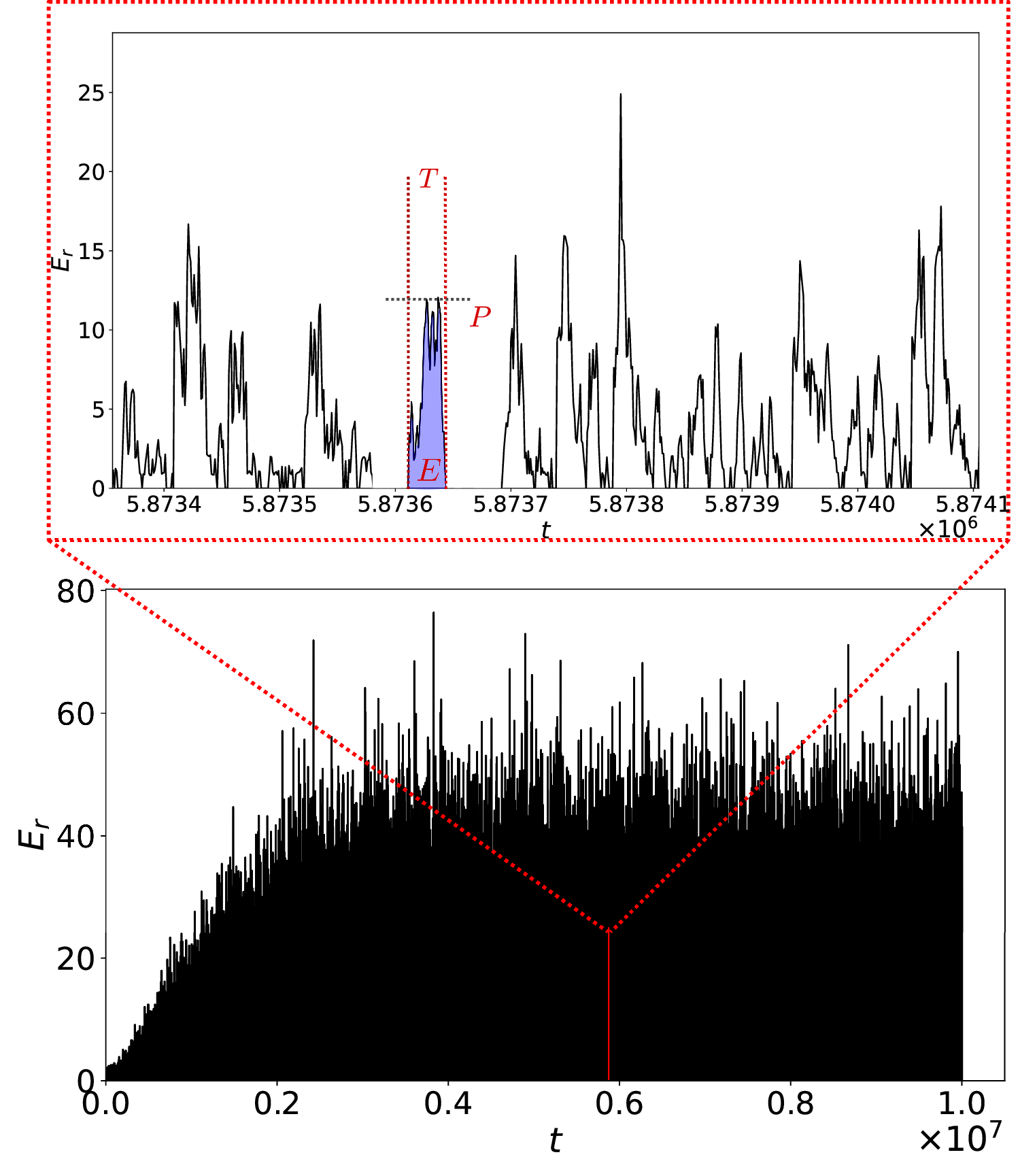}
		\caption{Dissipated energy from a 80-20 system. The red region in the lower panel shows a particular avalanche with total energy dissipated $E$, peak energy $P$, and duration $T$.}
		\label{fig:dissipatedenergypluszoom}
	\end{figure}

	An example of the dissipated energy $E_r$ as a function of time is shown in Fig.~\ref{fig:dissipatedenergypluszoom}, when $p_d=0.8$. A small section is zoomed out to explicitly show the existence of avalanches at short time scales, with a certain duration $T$, peak dissipated energy $P$, and dissipated energy during the event $E$.

\begin{figure}[ht!]
		\centering
		\begin{subfigure}[b]{0.23\textwidth}
			\includegraphics[width=\linewidth]{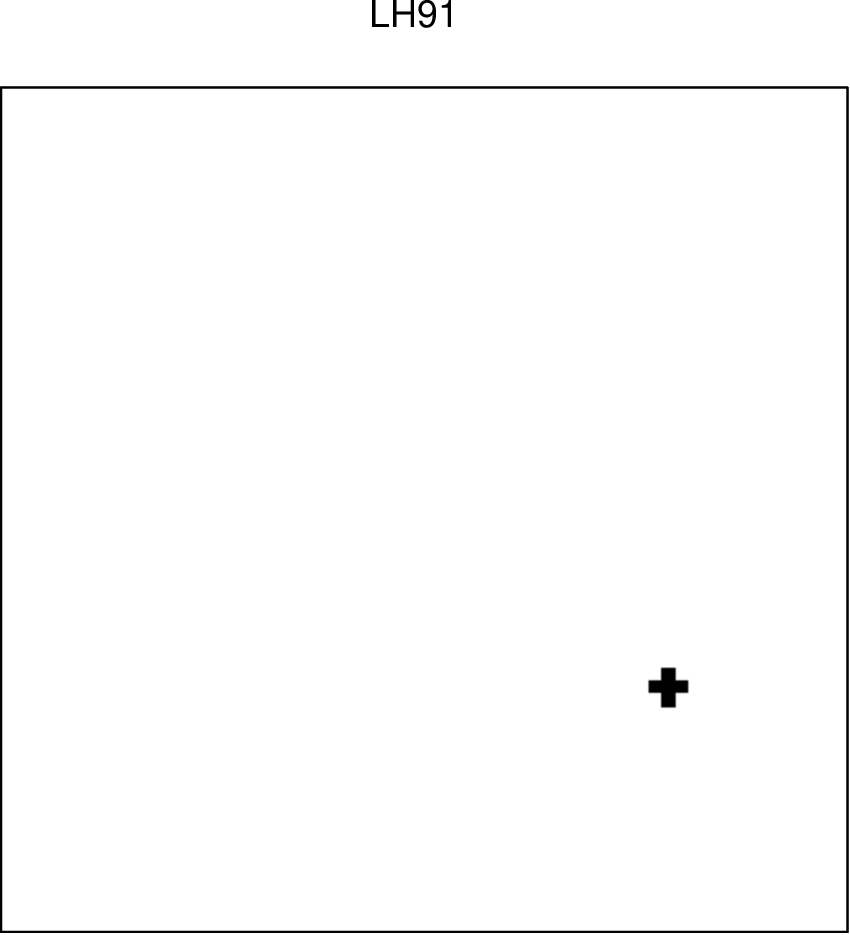}
			\caption{$t=1$}
			\label{fig:t0131}
		\end{subfigure}
		\hfill
		\begin{subfigure}[b]{0.23\textwidth}
			\includegraphics[width=\linewidth]{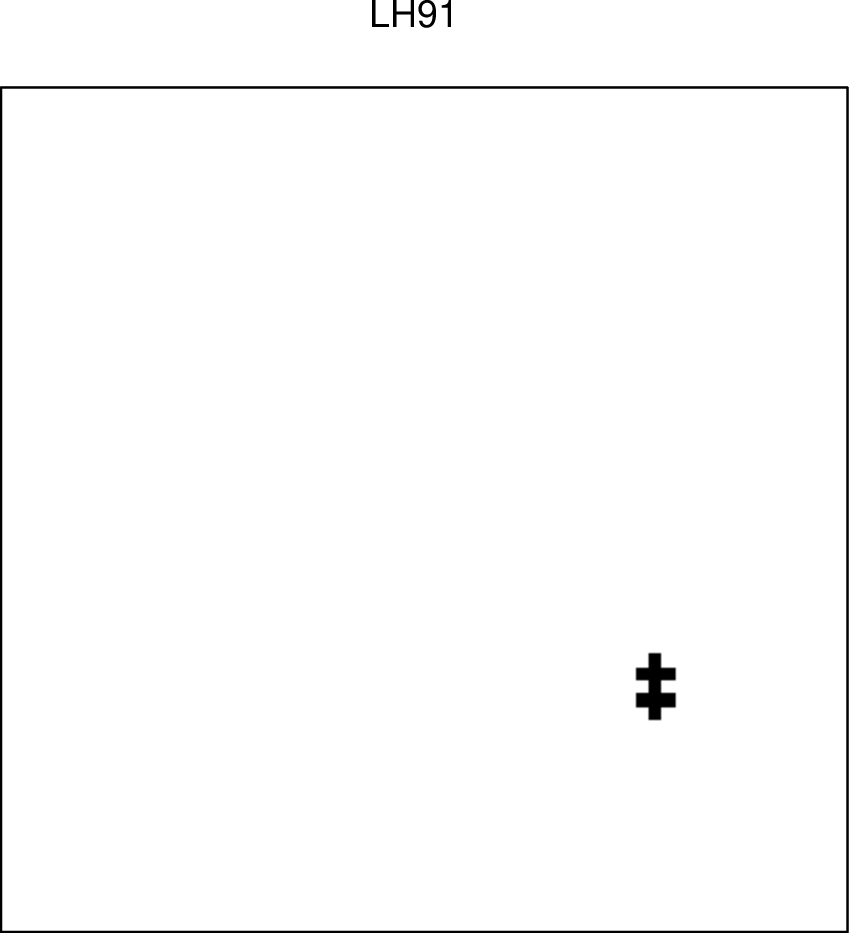}
			\caption{$t=3$}
			\label{fig:t2133}
		\end{subfigure}
		\hfill
		\begin{subfigure}[b]{0.23\textwidth}
			\includegraphics[width=\linewidth]{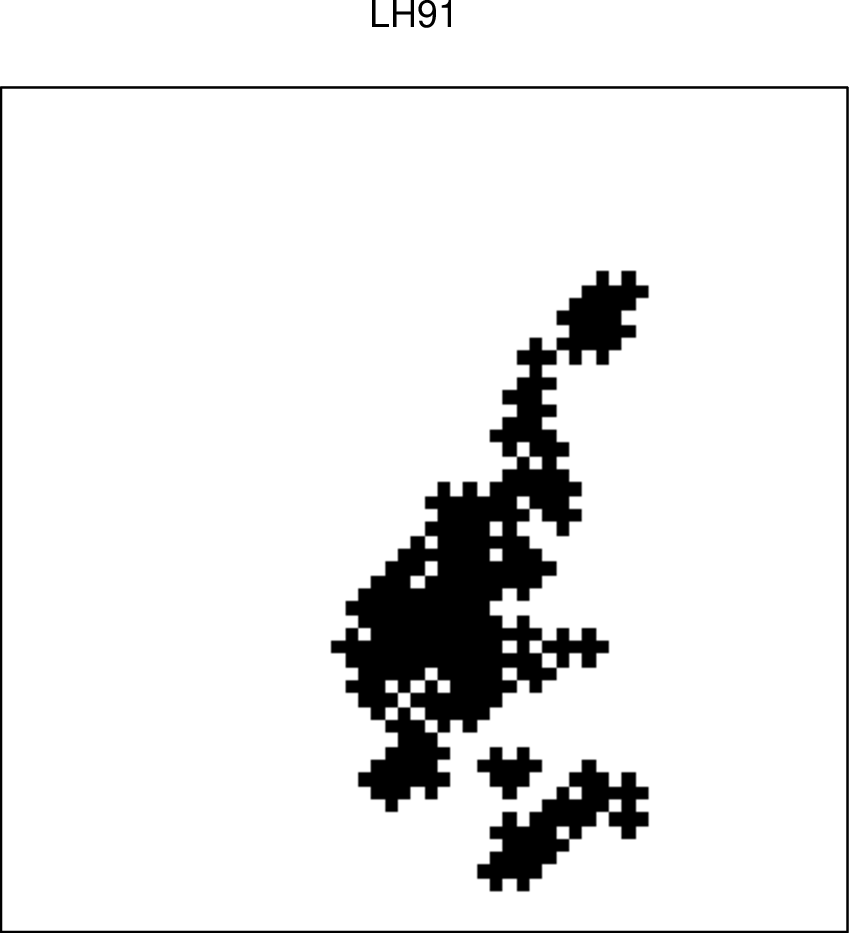}
			\caption{$t=t_{\text{peak}}$}
			\label{fig:tp300}
		\end{subfigure}
		\hfill
		\begin{subfigure}[b]{0.23\textwidth}
			\includegraphics[width=\linewidth]{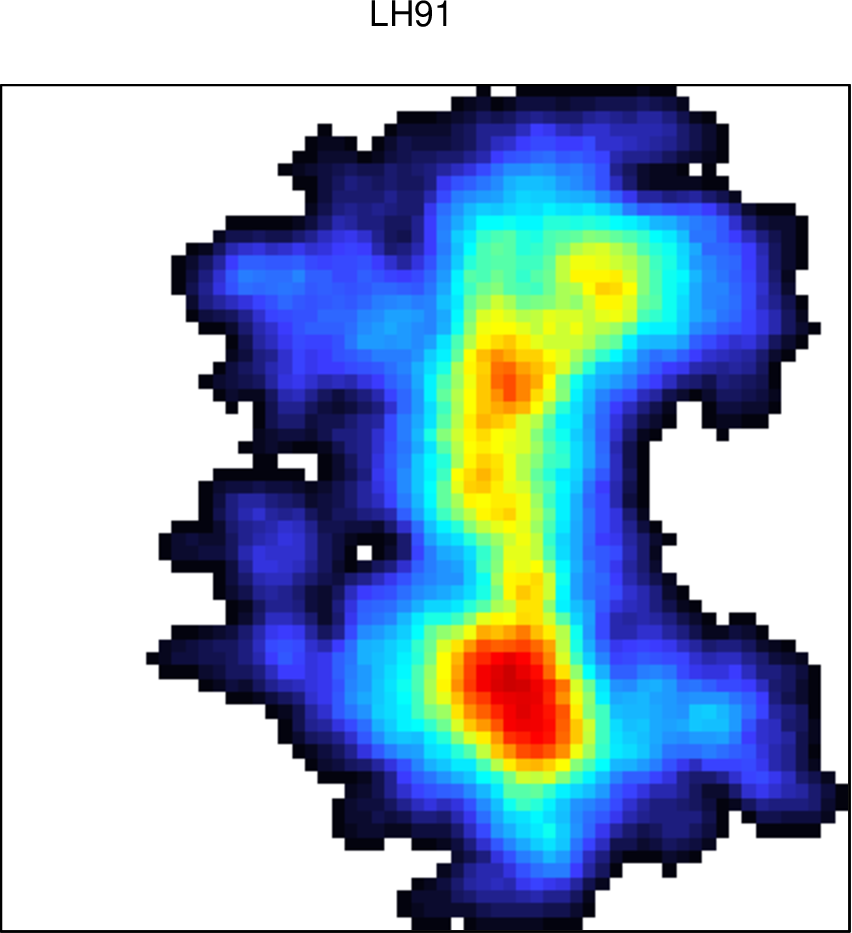}
			\caption{Total}
			\label{fig:t1132}
		\end{subfigure}
		
		\vspace{0.5cm}
		
		\begin{subfigure}[b]{0.23\textwidth}
			\includegraphics[width=\linewidth]{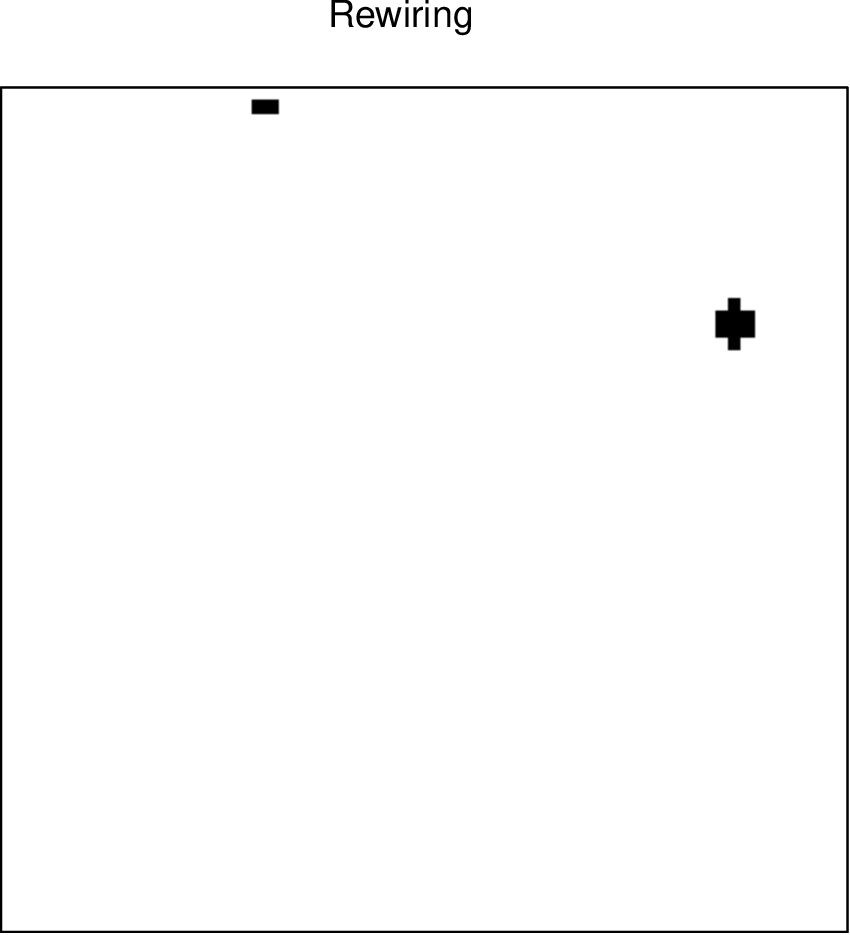}
			\caption{$t=1$}
			\label{fig:t09064}
		\end{subfigure}
		\hfill
		\begin{subfigure}[b]{0.23\textwidth}
			\includegraphics[width=\linewidth]{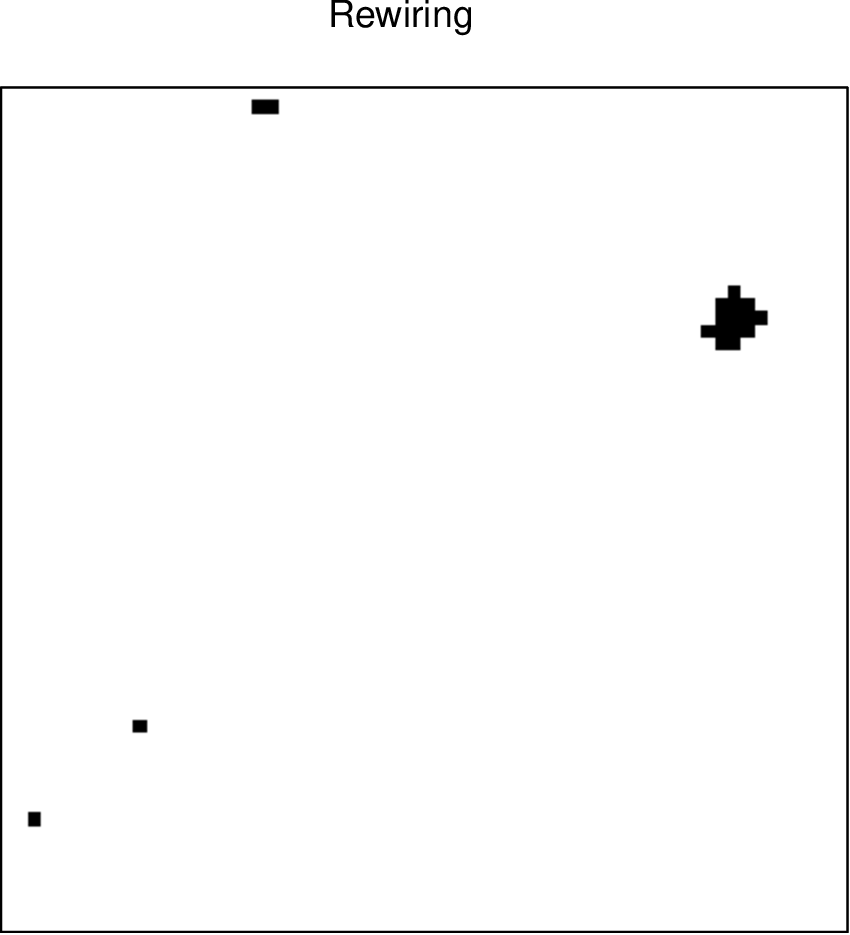}
			\caption{$t=3$}
			\label{fig:t29066}
		\end{subfigure}
		\hfill
		\begin{subfigure}[b]{0.23\textwidth}
			\includegraphics[width=\linewidth]{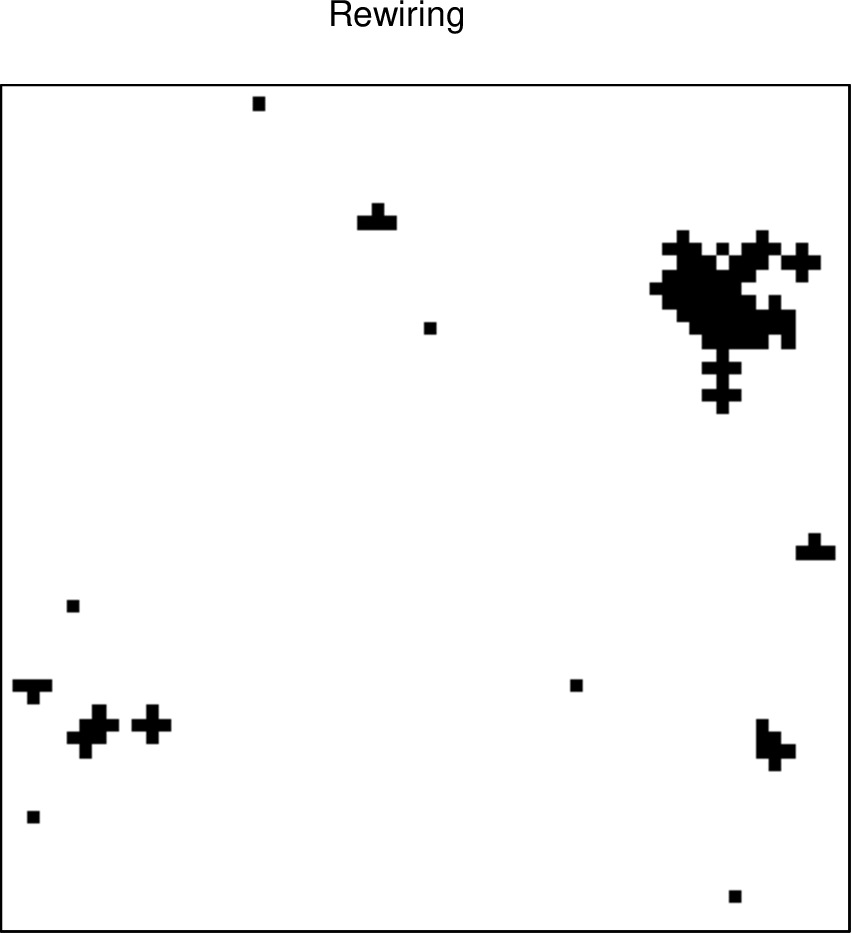}
			\caption{$t=t_{\text{peak}}$}
			\label{fig:tpeak9096}
		\end{subfigure}
		\hfill
		\begin{subfigure}[b]{0.23\textwidth}
			\includegraphics[width=\linewidth]{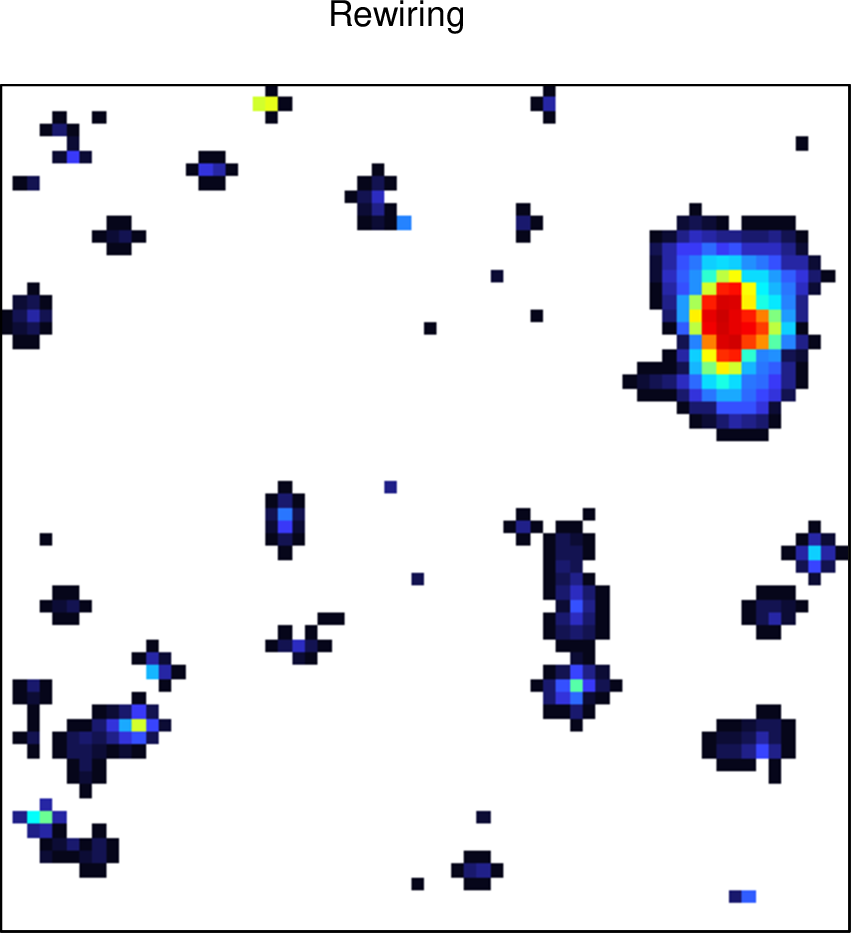}
			\caption{Total}
			\label{fig:t19065}
		\end{subfigure}
		
		\caption{Spatial distribution of one avalanche: the top row shows the original model without rewiring, and the bottom row corresponds to an 80-20 rewiring configuration. Panels (a) and (e) depict the beginning of the avalanche; (b) and (f) show the system at time $t_3$; (c) and (g) show the state at the peak of the avalanche; and (d) and (h) show the full extent of the avalanche-affected area, with impact frequency visualized on a color scale: white (0 times affected), black (1 time), and a rainbow gradient for larger values.}
		\label{fig:comparacion}
	\end{figure}

	Figure~\ref{fig:comparacion}, on the other hand, shows the spatial evolution of one avalanche, which allows to compare the case of the LH91 model (upper panel) and the rewired model (lower panel). It is clear that, for the rewired grid, the avalanche process does not start in a single neighborhood, showing that the avalanche condition and the flow of released energy involved topological rather than geometrical neighbors. For the same reason, when the avalanche reaches peak activity, instead of a single, connected region, the avalanche involves site in unconnected (in the geometrical sense) islands, but topologically connected due to rewiring.
	
\begin{figure*}[hp]
		\centering
		% --- First row---
		\begin{subfigure}[t]{0.32\textwidth}
			\includegraphics[width=\linewidth]{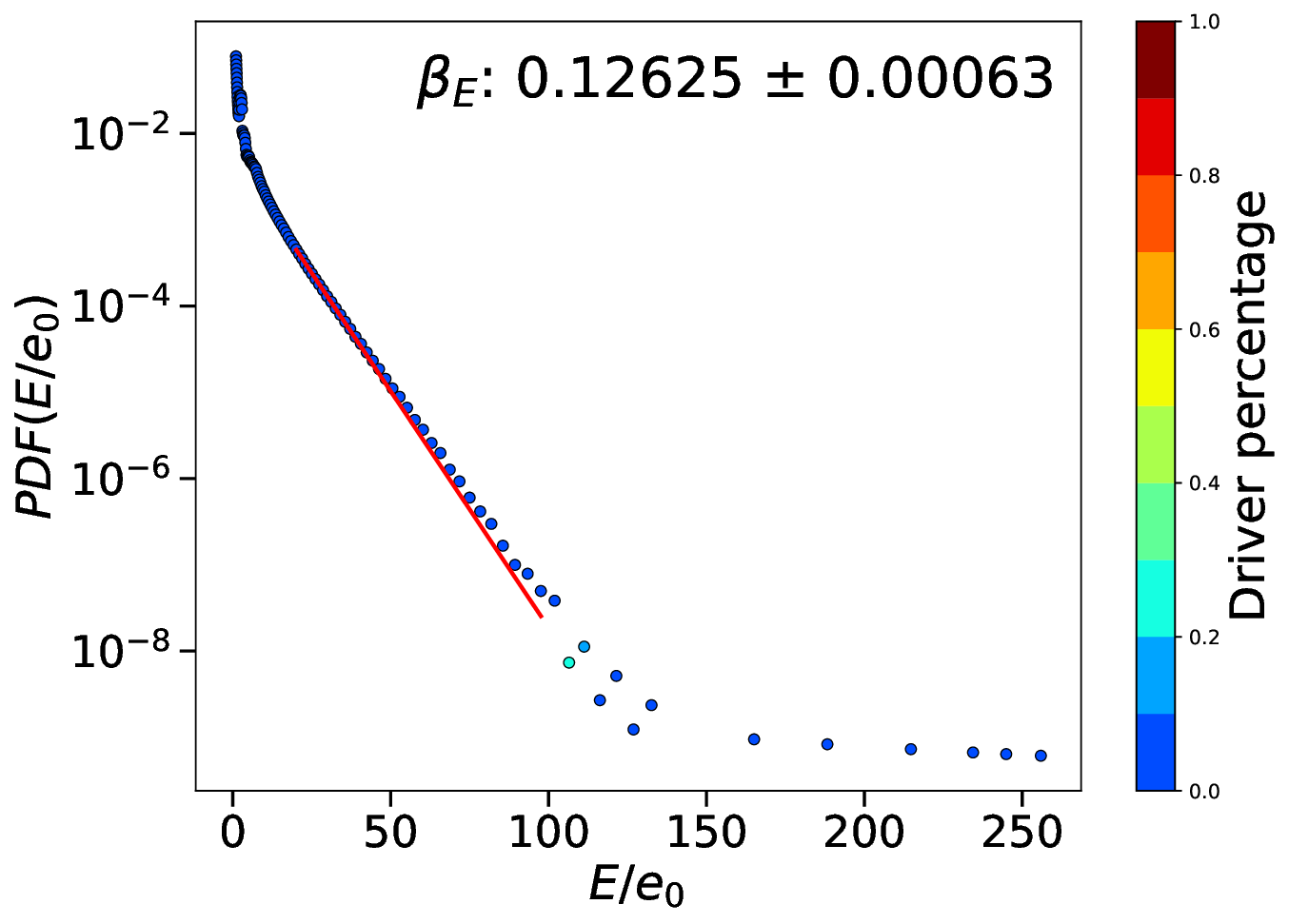}
			\caption{Driving-Rewiring: 10-90}
			\label{fig:pdf10}
		\end{subfigure}
		\hfill
		\begin{subfigure}[t]{0.32\textwidth}
			\includegraphics[width=\linewidth]{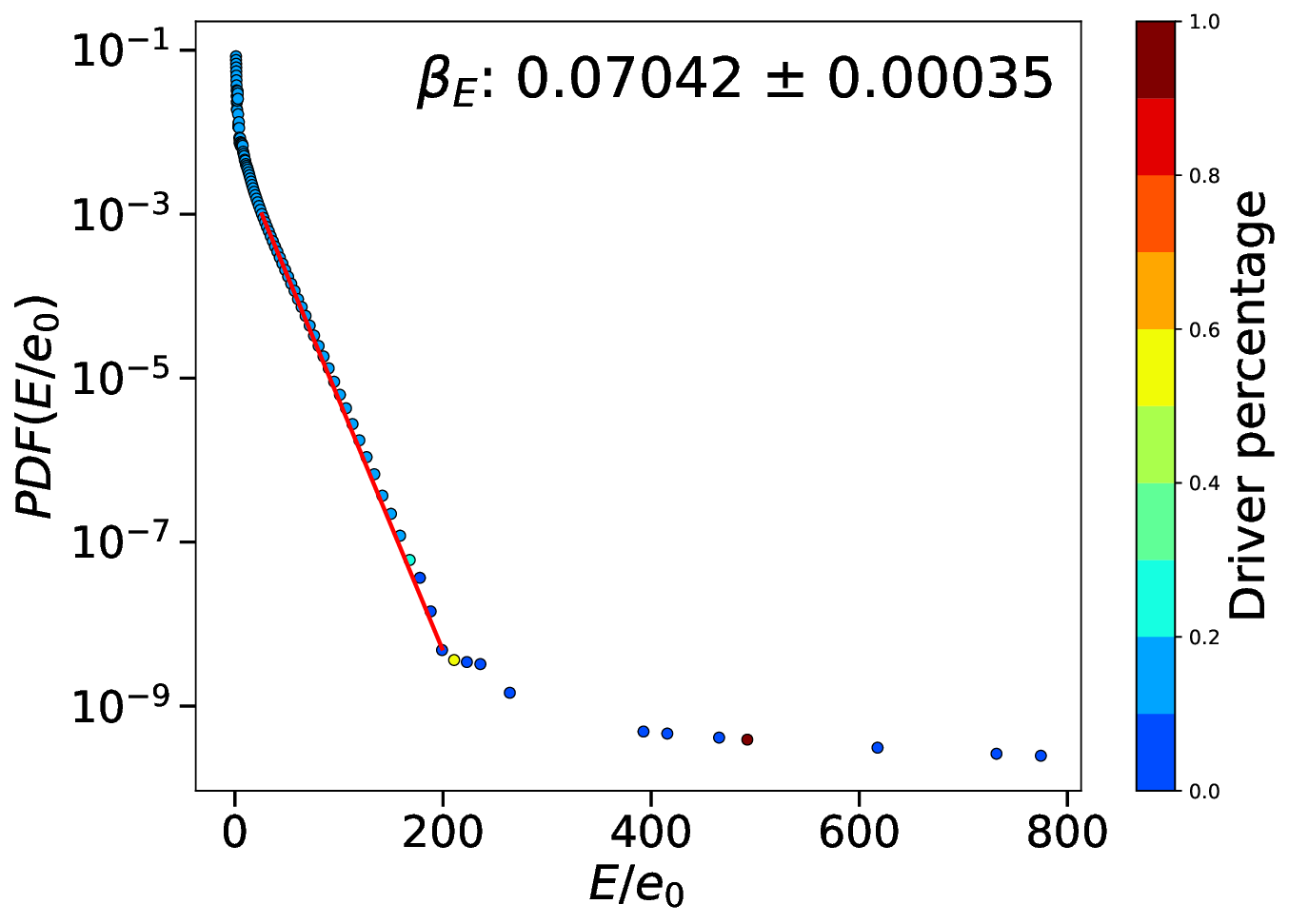}
			\caption{Driving-Rewiring: 20-80}
			\label{fig:pdf20}
		\end{subfigure}
		\hfill
		\begin{subfigure}[t]{0.32\textwidth}
			\includegraphics[width=\linewidth]{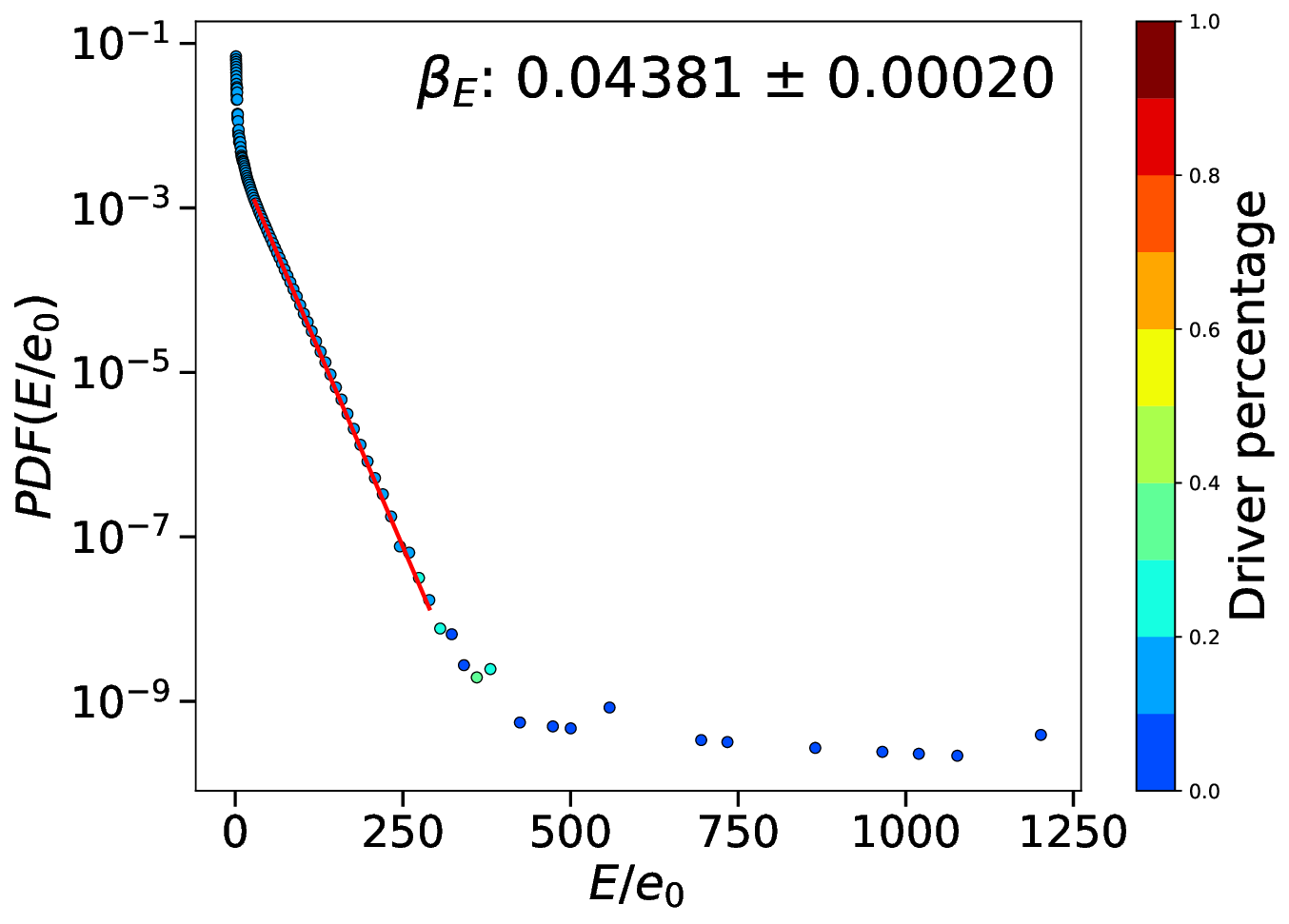}
			\caption{Driving-Rewiring: 30-70}
			\label{fig:pdf30}
		\end{subfigure}
		
		% --- Second row---
		\begin{subfigure}[t]{0.32\textwidth}
			\includegraphics[width=\linewidth]{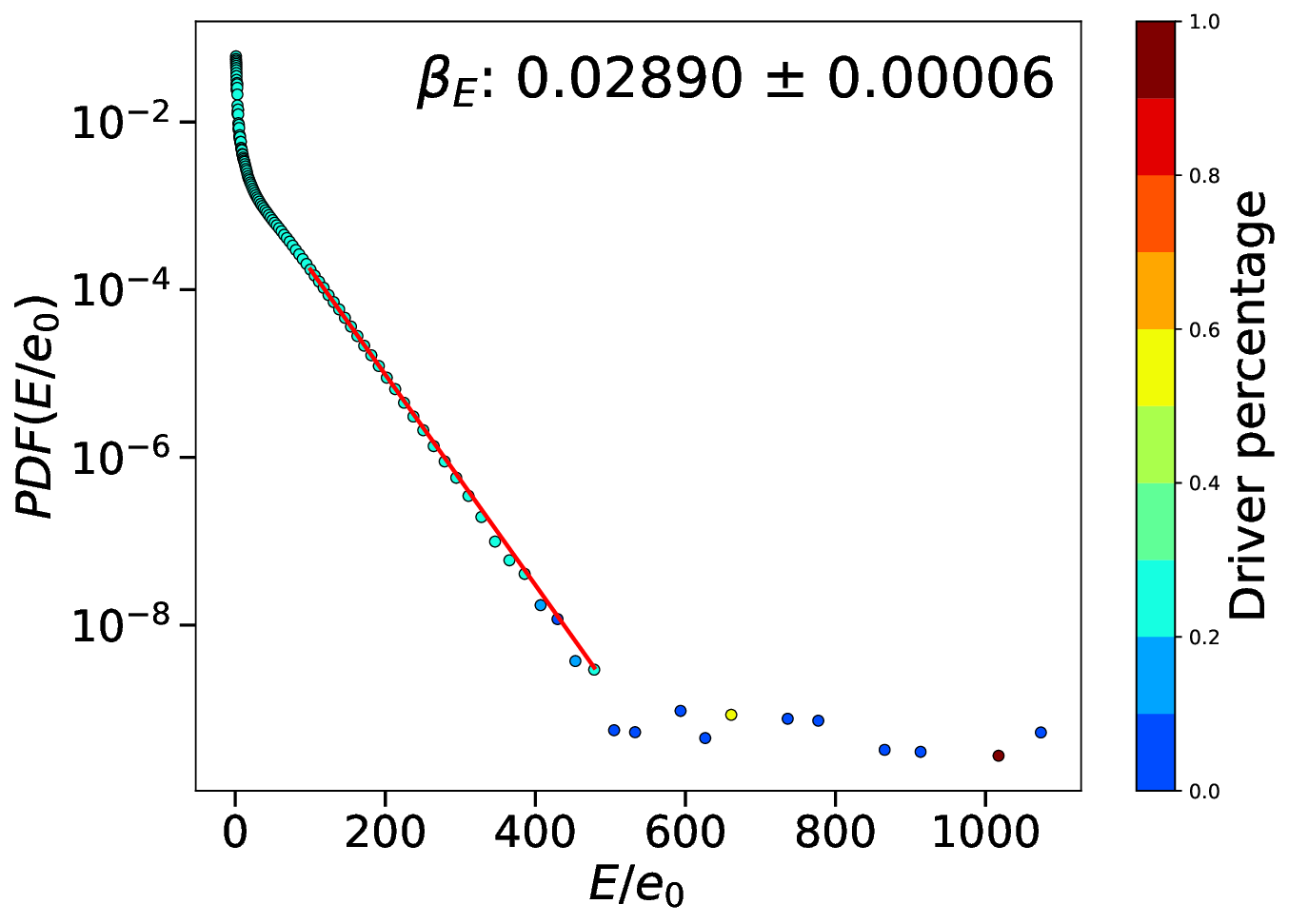}
			\caption{Driving-Rewiring: 40-60}
			\label{fig:pdf40}
		\end{subfigure}
		\hfill
		\begin{subfigure}[t]{0.32\textwidth}
			\includegraphics[width=\linewidth]{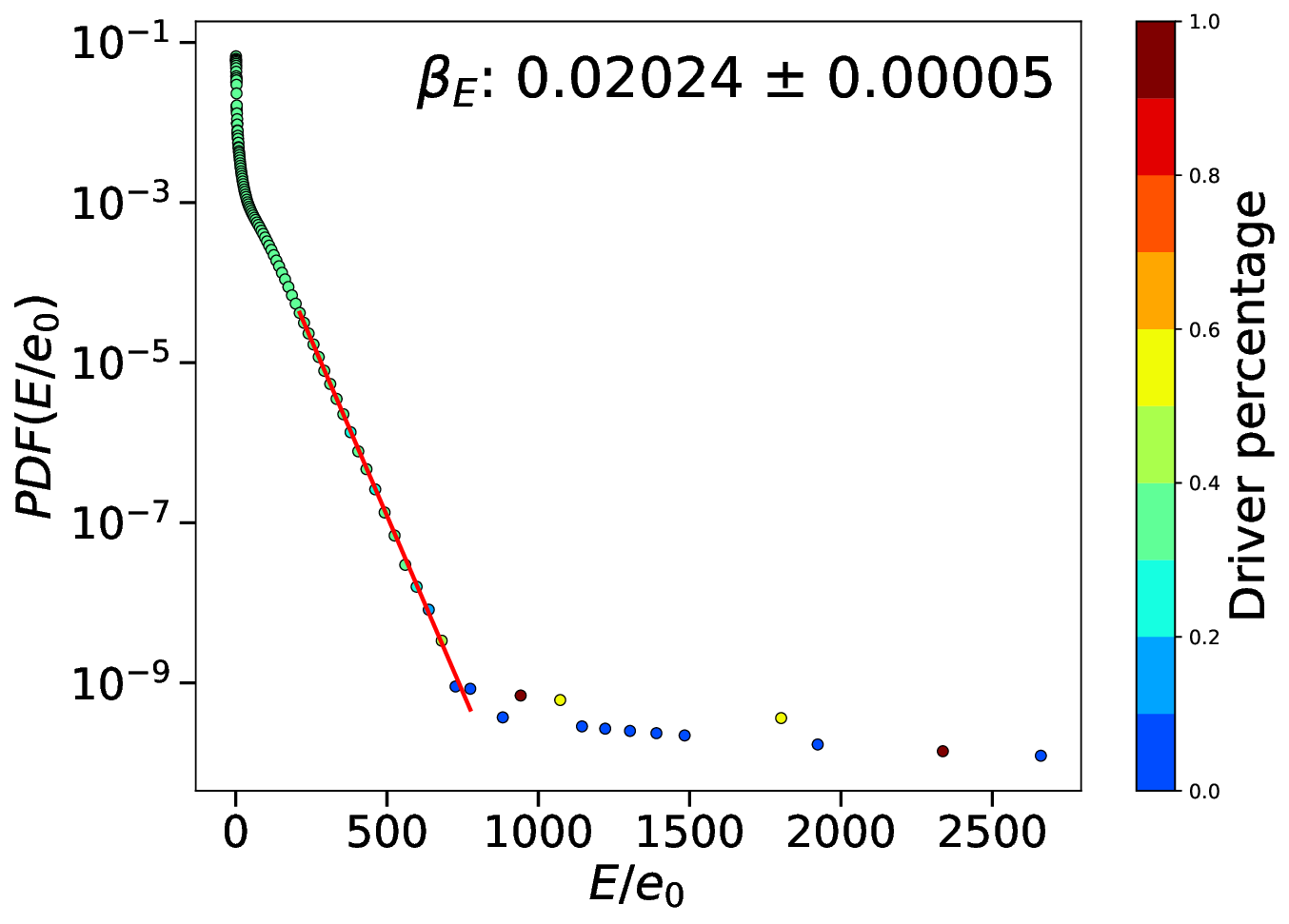}
			\caption{Driving-Rewiring 50-50}
			\label{fig:pdf50}
		\end{subfigure}
		\hfill
		\begin{subfigure}[t]{0.32\textwidth}
			\includegraphics[width=\linewidth]{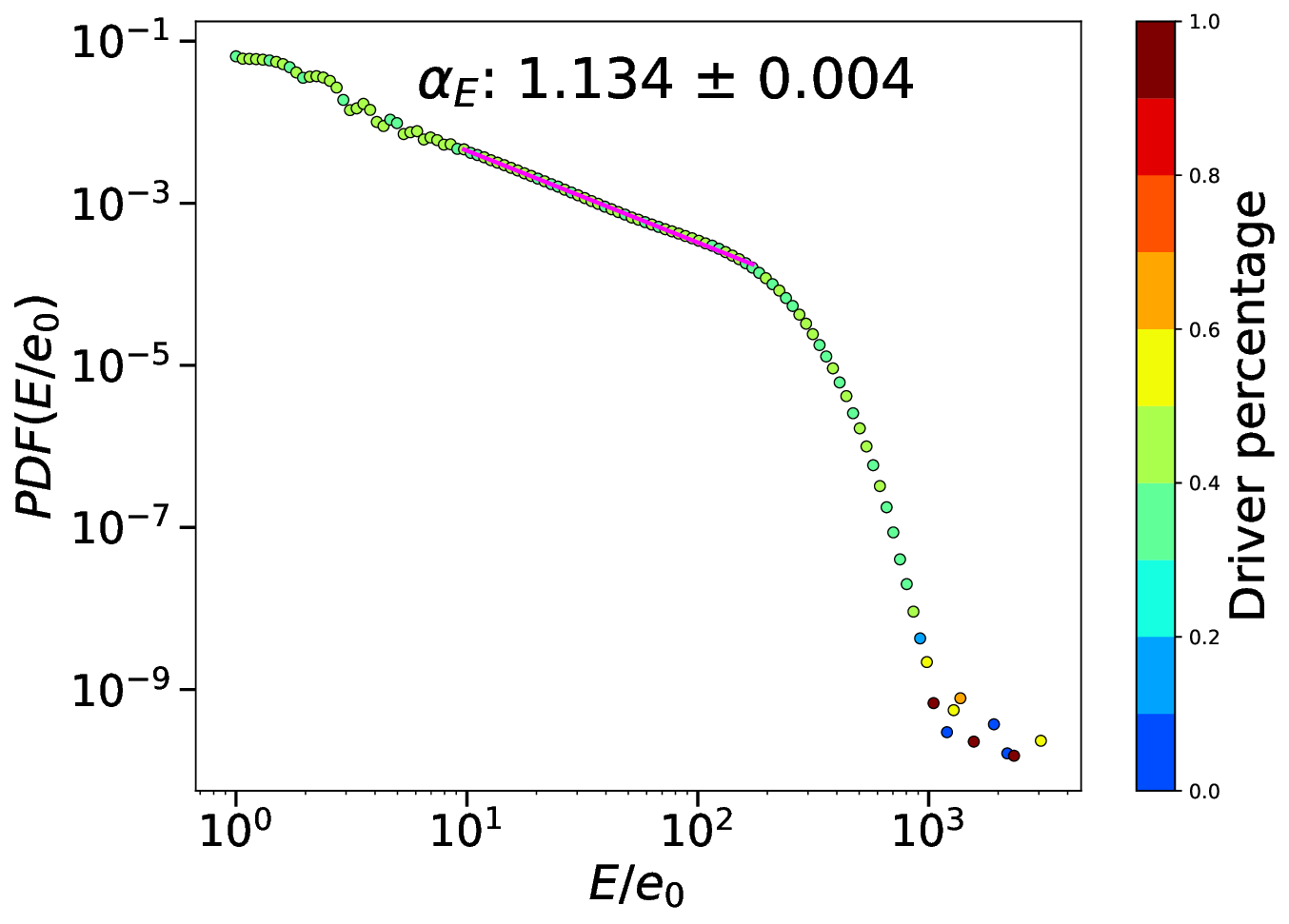}
			\caption{Driving-Rewiring: 60-40}
			\label{fig:pdf60}
		\end{subfigure}
		
		% --- Third row---
		\begin{subfigure}[t]{0.32\textwidth}
			\includegraphics[width=\linewidth]{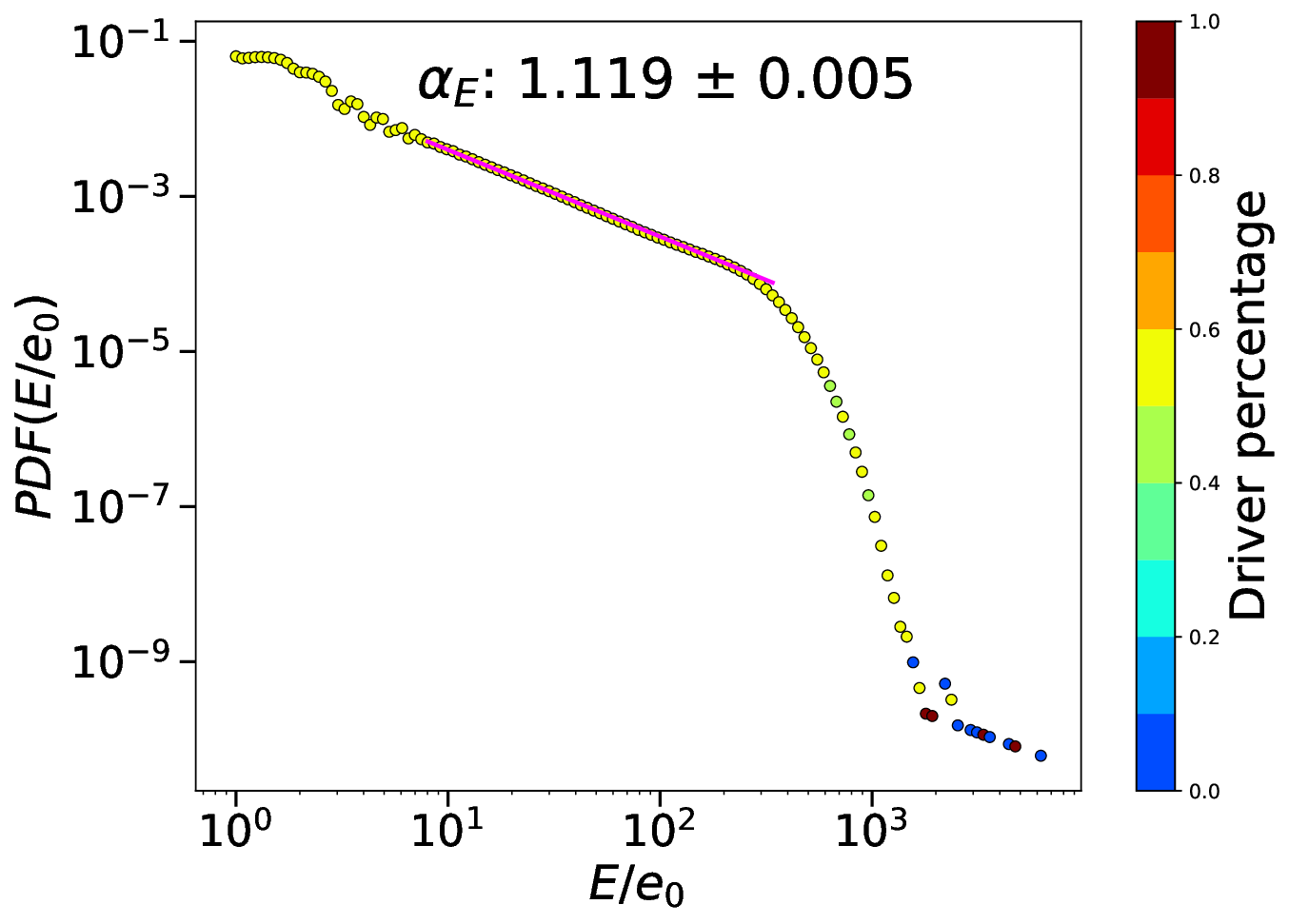}
			\caption{Driving-Rewiring: 70-30}
			\label{fig:pdf70}
		\end{subfigure}
		\hfill
		\begin{subfigure}[t]{0.32\textwidth}
			\includegraphics[width=\linewidth]{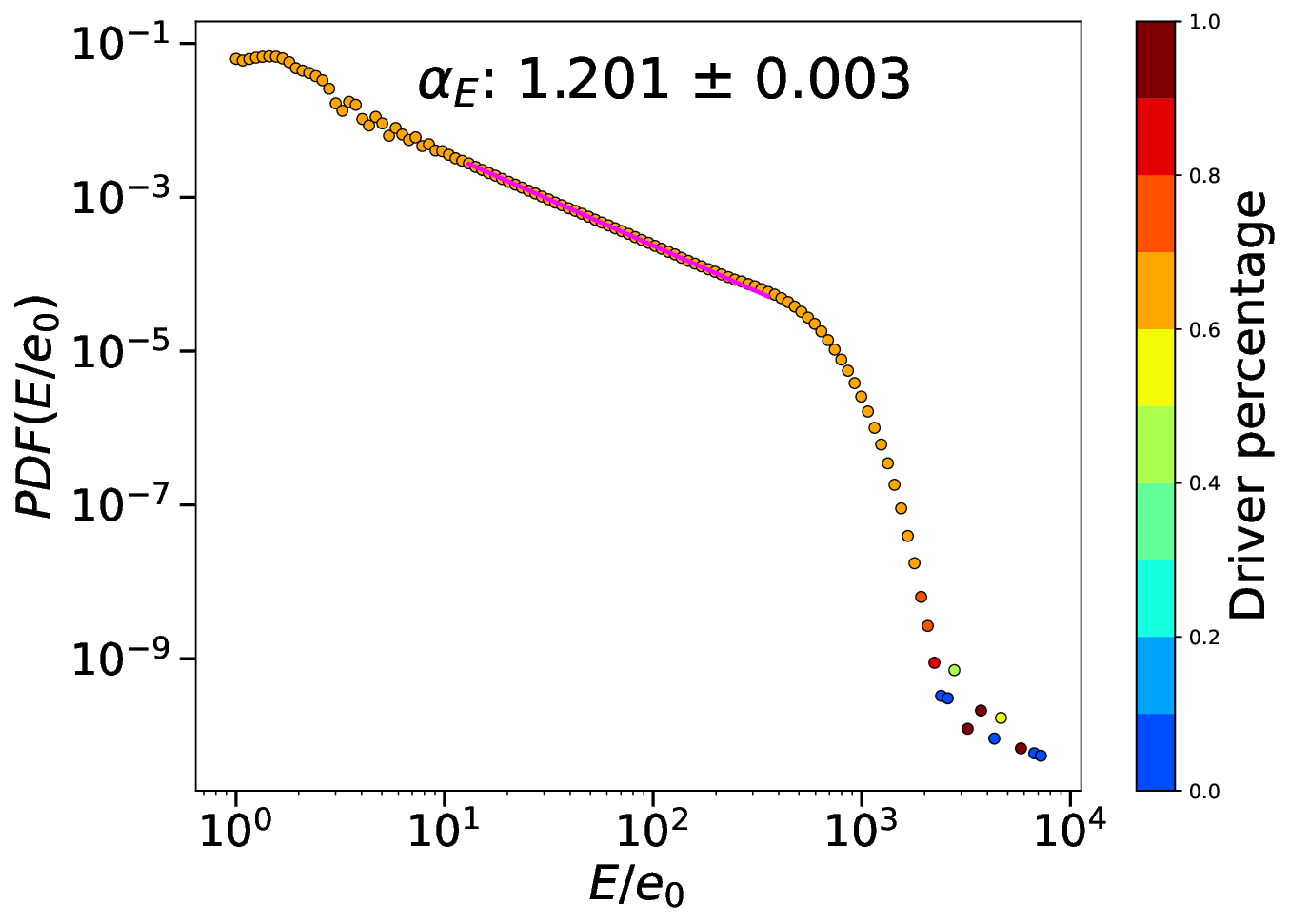}
			\caption{Driving-Rewiring: 80-20}
			\label{fig:pdf80}
		\end{subfigure}
		\hfill
		\begin{subfigure}[t]{0.32\textwidth}
			\includegraphics[width=\linewidth]{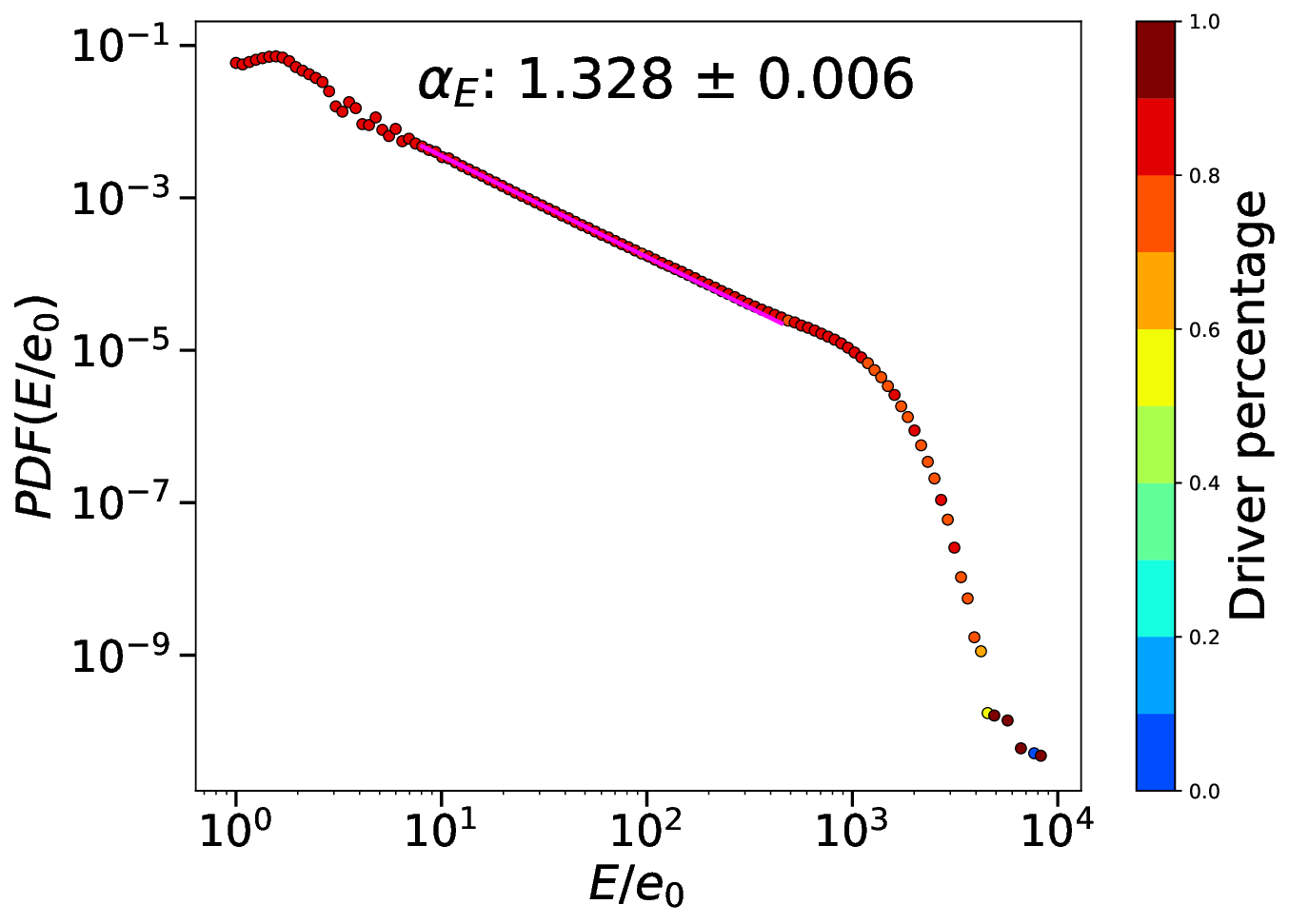}
			\caption{Driving-Rewiring: 90-10}
			\label{fig:pdf90}
		\end{subfigure}
		
		\caption{PDF colored by probability for Driving-Rewiring values from 10 to 90. Each subplot shows the distribution for a specific Driving-Rewiring. Using $10^7$ avalanches, $N=64$. The color bar represents what is the percentage of events considered in each bin that was produced by driving gain, $0\%$ implies events produced entirely by rewiring, while $100\%$ implies events produced entirely by driving.}
		\label{fig:pdf_grid_3x3}
	\end{figure*}

	Using this new model, we compute the probability distribution function (PDF) of the energy released across all simulations performed. This is shown in Fig.~\ref{fig:pdf_grid_3x3}, where the distributions obtained for different values of the rewiring probability are shown. We observe that, if rewiring is the dominant effect [Figs.~\ref{fig:pdf_grid_3x3}(a)--(d)], the distribution decays exponentially for larger energies, whereas if the external driver is more important, it exhibits a power-law behavior. Thus, in the LH91 limit, where no rewiring occurs, the expected scale-free behavior is recovered.
	
	As mentioned in the previous section, avalanche events can be triggered either by the external driver or by rewiring, which modifies the neighbour configuration, thus it is interesting to see how these two types of avalanche contribute to the final distribution. This is shown in Fig.~\ref{fig:pdf_grid_3x3} by coloring the dots. Every dot represents the number of avalanche events for a given energy, and its color represents the proportion of those events which were triggered by the driving mechanism. Green/blue dots mean that most avalanches for that energy were triggered by rewiring, and orange/red dots mean that they were essentially triggered by driving.   All simulations begin with the original grid configuration (no initial rewiring), and each distribution is obtained from $10^7$ avalanches. 
	
	It can be observed that, as the driver/rewiring ratio is modified, both the nature of the distribution (exponential/scale free), and the decay exponents change.

	%We are now able to review the energy dissipated in the new model. For this, we plot the probability of dissipated energy versus the normalized dissipated energy and we apply a least-squares fit to the best-fit curve to obtain the function that best fits the data.
	%Additionally, in Fig.~\ref{fig:pdf_grid_3x3} we analyze the frequency of avalanches triggered by two distinct mechanisms: (i) \emph{driving}, which, as in classical models, occurs when the magnetic field exceeds the critical threshold $Z_c $, and (ii) \emph{rewiring}, which modifies the topological neighborhood, allowing the system to surpass $Z_c$ through new connections. 

	%For the rewiring model described here, we analyze the statistics of 
	%energy distributions across different rewiring probability configurations in Fig.~\ref{fig:pdf_grid_3x3}, ranging from 10-90\% to 90-10\%. Each distribution was constructed from $10^7$ avalanches, revealing distinct critical exponents (see Fig.~\ref{fig:pdf_grid_3x3}). All simulations began with the original grid configuration (no initial rewiring), with each rewiring event permanently altering the network until subsequent rewiring modified it further. 
	
	In order to assess the change from exponential to scale-free behavior in a more systematic manner, we impose an exponential and a power-law fit to each distribution, and use the relative error of the exponent resulting from the fit as an estimation of the quality of each representation. If the distribution is exponential, one would expect a small relative error for the decay exponent of the exponential fit; and if the distribution is scale-free, the error should be smaller for the power-law fit. A similar idea has been previously used to study transition from exponential to scale-free behavior in complex networks models~\cite{guillier2017optimization}. 
	
\begin{figure*}[ht!]
		\centering
		\begin{subfigure}[b]{0.49\textwidth}
			\centering
			\includegraphics[width=\linewidth]{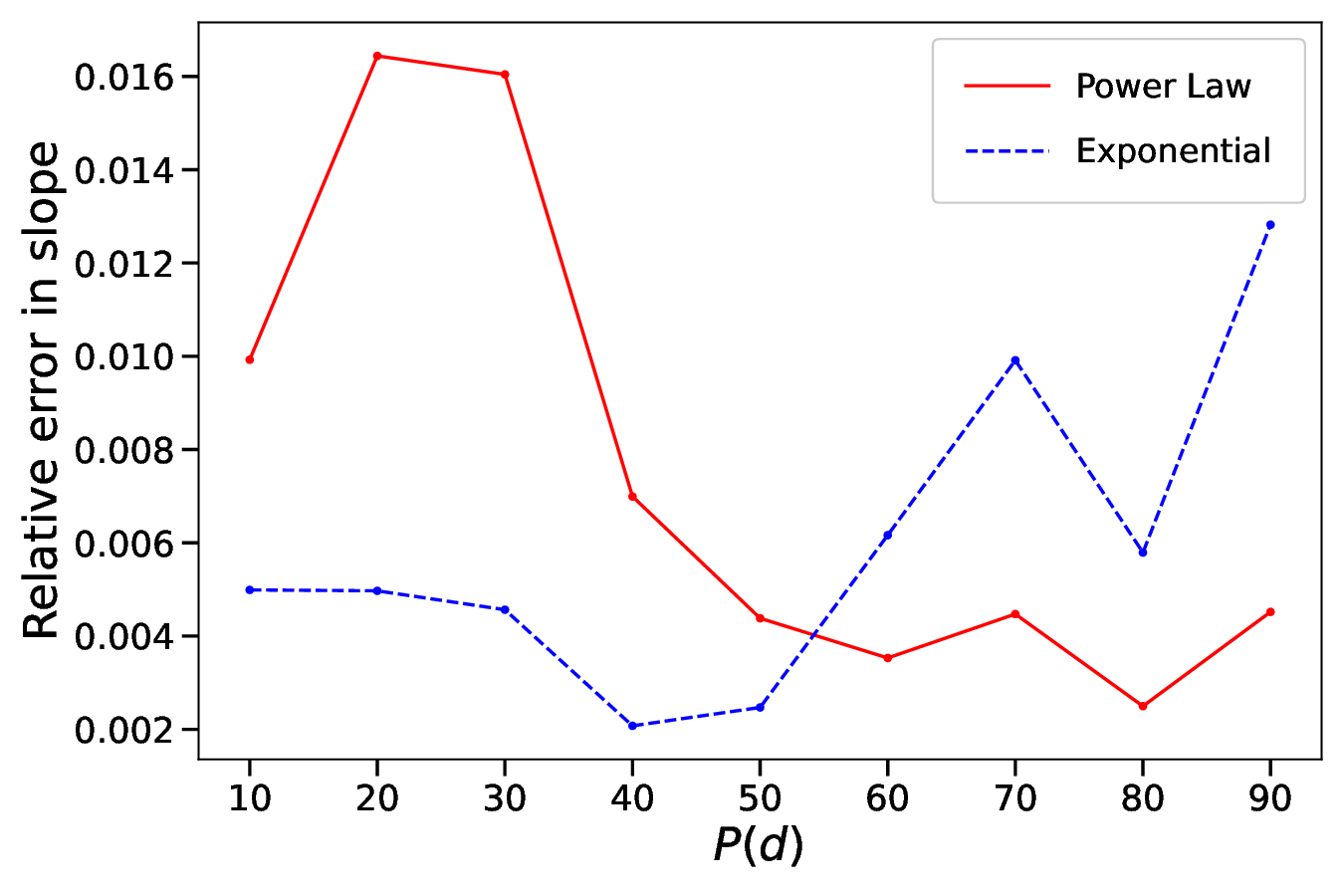}
			\caption{}
			\label{fig:relative_error_p}
		\end{subfigure}
		\hfill
		\begin{subfigure}[b]{0.49\textwidth}
			\centering
			\includegraphics[width=\linewidth]{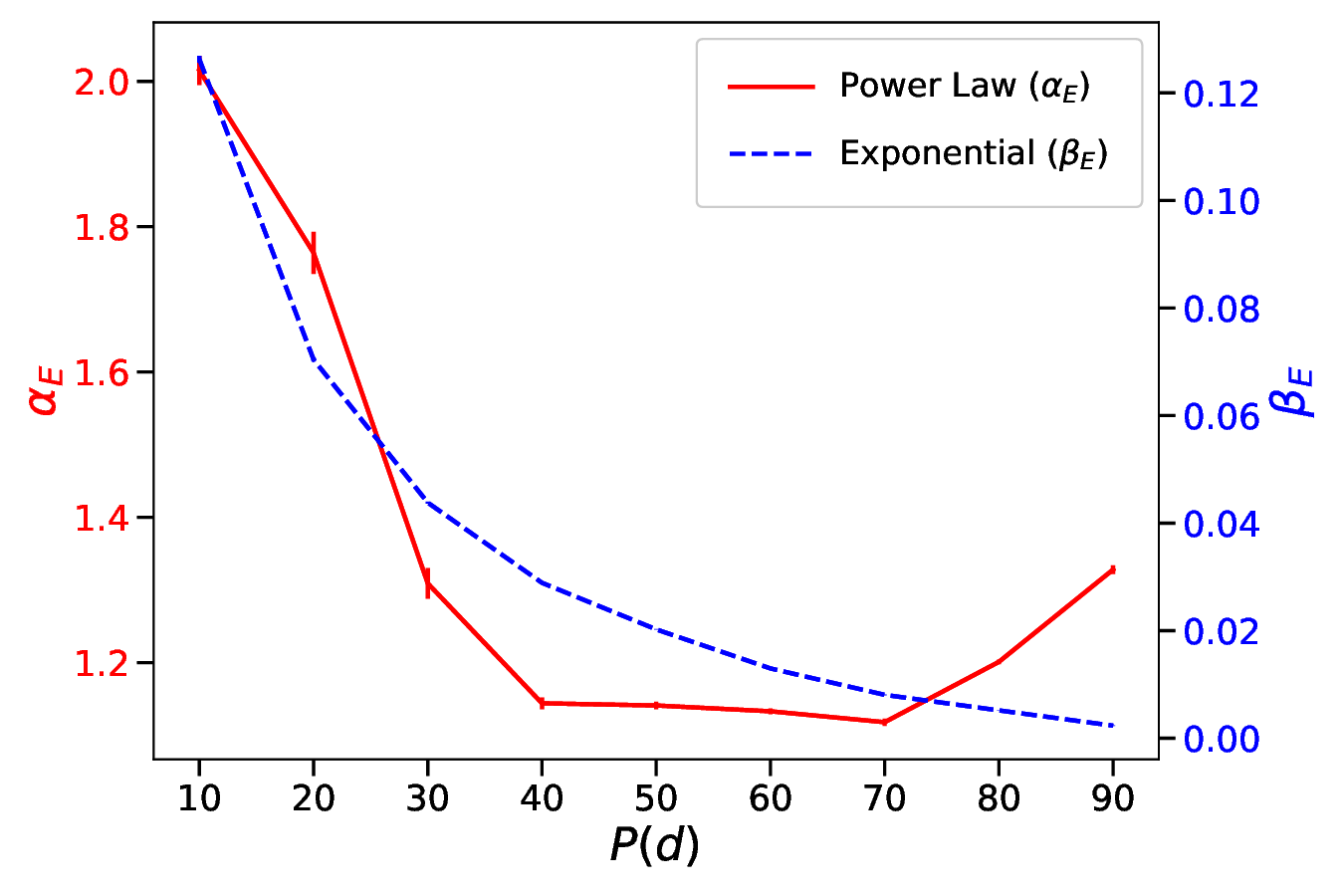}
			\caption{}
			\label{fig:powerlaw_exponential}
		\end{subfigure}
		\caption{(a) Relative error for the best fit of the decay exponent for a power-law model (red) or exponential model (blue), for the dissipated energy distribution, as a function of the probability of driving $p(d)$ rather than rewiring. (b) Decay exponents obtained for each model: power law ($\alpha_E$, red) and exponential ($\beta_E$, blue), model, as a function of $p(d)$.}
		\label{fig:powerlawvsexponentialtwinx}
	\end{figure*}

	This is shown in Fig.~\ref{fig:powerlawvsexponentialtwinx}(a). As expected, the exponential fit is better when rewiring dominates, and the power-law fit is better when driving dominates. The transition from one behavior to the other occurs approximately when driving starts to dominate, at $p(d)\sim 0.55$,  which justifies the choices for the fits in Fig.~\ref{fig:pdf_grid_3x3}. 
	
	On the other hand, Fig.~\ref{fig:powerlawvsexponentialtwinx}(b) shows all values for the best fits for the exponents, regardless of whether the best fit is power law or exponential. This allows to follow the trend for each decay exponent, in the regions where they are more reliable.
	
	Results in Figs.~\ref{fig:powerlawvsexponentialtwinx} are summarized as a table in Table~\ref{table_exponentes}, which shows the best fits for the power law exponent $\alpha_E$, the exponential decay exponent $\beta_E$, and their respective errors. 
	
	\begin{table*}[ht!]
		\centering
		\caption{Comparison of power law and exponential decay exponent values and errors.}
		\label{table_exponentes}
		\begin{tabular}{|c|c|c|c|c|}
			\hline
			{Driving-Rewiring \%} &  {$\alpha_E$} {(Power Law)} &  {Error $\alpha_E$} &  {$\beta_E$ (Exponential)} &  {Error $\beta_E$} \\
			\hline
			10-90 & 2.015 & $\pm$0.020 & 0.12625 & $\pm$0.00063 \\
			20-80 & 1.764 & $\pm$0.029 & 0.07042 & $\pm$0.00035 \\
			30-70 & 1.309 & $\pm$0.021 & 0.04381 & $\pm$0.00020 \\
			40-60 & 1.144 & $\pm$0.008 & 0.02890 & $\pm$0.00006 \\
			50-50 & 1.141 & $\pm$0.005 & 0.02024 & $\pm$0.00005 \\
			60-40 & 1.134 & $\pm$0.004 & 0.01298 & $\pm$0.00008 \\
			70-30 & 1.119 & $\pm$0.005 & 0.00807 & $\pm$0.00008 \\
			80-20 & 1.201 & $\pm$0.003 & 0.00518 & $\pm$0.00003 \\
			90-10 & 1.328 & $\pm$0.006 & 0.00238 & $\pm$0.00003 \\
			\hline
		\end{tabular}
	\end{table*}

	\section{Summary and Discussion}
	\label{summary}
	
	By framing energy transport in terms of a complex network, we have reformulated the LH model as a network-based transport process, where magnetic field dynamics are governed by topological neighbor interactions. The resulting energy dissipation in this system exhibits a power-law distribution, consistent with avalanche-like behavior observed in solar flare statistics \cite{1993AdSpR..13..179C,aschwanden2025scaling}. Interestingly, the power-law exponents ($\alpha_E$) in our model fall within the range of 1.07--1.39, aligning with values reported in other studies~\cite{2014SoPh..289.4137S}. These models propose a nonlocal dynamics, which is also the case for the model presented here, when rewiring is introduced in the LH91 model. Nonlocality leads to avalanches occurring in different regions, preventing the system from accumulating excessive energy. Instabilities at various points facilitate energy release, maintaining a balanced state. The global dynamics are illustrated in Fig.~\ref{fig:comparacion}, which also provides a comparison with the original model, where interactions are strictly local due to nearest-neighbor coupling.
	
	This raises important questions for solar flare modeling: Should rewiring effects be incorporated? If so, how? In our implementation, we balance rewiring and driving through mixed probabilities, finding that rewiring typically dominates flare generation and leads to shorter waiting times between events visible in Fig.~\ref{fig:dissipatedenergypluszoom}. Notably, when both processes occur simultaneously (rewiring and driving), the dynamics remain similar to the 50-50 case. While our mixed system explores a broader parameter space, it still falls short of reproducing realistic solar flare distributions.
	
	The transition between rewiring-dominated and driving-dominated regimes significantly impacts energy distribution exponent and production of avalanches, with rewiring producing more frequent, lower-energy flares, as is shown in Fig.~\ref{fig:pdf_grid_3x3}.  Variations in the power law exponents for waiting times between events have been studied in~\cite{baiesi2006intensity}, obtaining $\gamma_w^{+} = 1.51$ for solar maximum and $\gamma_w
	^{-}=2.83$ for solar minimum. Since waiting times are correlated with both energy peaks and magnetic energy released by the system~\cite{2001SoPh..203..321C}, we can argue that the variations we observe in the value of $\alpha_E$ may be associated with different stages of the solar cycle, with lower values during solar maximum, and larger values during solar minimum. {This result is consistent with previous works showing
		that the decay exponent of flare energy distribution depends on
		the phase of the solar cycle,\citep{aschwanden2012automated,biasiotti2025statistical}. In particular, and restricting} the discussion to the region where the power law fit is better [$p_d>0.5$ in Fig.~\ref{fig:powerlawvsexponentialtwinx}(a)], this suggests that the largest value, $\alpha_E=1.328$, signals a relationship between solar minimum and dominance of driver over rewiring events; and the smallest value, $\alpha_E=1.134$, suggests a relationship between solar maximum and an increased importance of rewiring events.
	
	The waiting times between consecutive avalanches is a topic that has been addressed in various ways within the framework of the classical LH model, notably through its canonical definition by \cite{wheatland2000origin}. More recently, \cite{kychenthal2023alternative} explored alternative definitions of waiting times for both the complete set of avalanches and for extreme events obtaining log-normal distributions in some cases. The considerations of our model suggest that it would be particularly interesting to extend this type of analysis to understand how waiting-time statistics are affected by rewiring. Since our model demonstrates that rewiring accelerates avalanche production and introduces nonlocal triggering, we hypothesize a significant impact on interevent statistics, potentially shifting them away from the established paradigms. A comprehensive analysis of this phenomenon, however, is beyond the scope of this paper, which focuses on energy distributions. Therefore, a detailed analysis of waiting-time statistics in our rewiring-based cellular automaton needs to be studied in another article.
	
	The framework presented here can be scaled naturally to larger systems, where interactions between multiple active regions (rather than isolated ones) become significant, which can be related to models where flares may influence each other~\cite{guite2025avalanching}, an issue which we plan to address in the future as well. 
	
	In summary, this model extends the LH91 framework to a complex network representation, naturally incorporating the physical evolution of magnetic field configurations and their activity variations as the energy release events progress. Rewiring accelerates avalanche production compared to the original model, as it introduces dissipation without injecting additional energy into the system, leading to non-scale-free avalanche statistics if rewiring dominates over driving. However, if rewiring is present, but does not dominate over driving, the power-law behavior is reproduced by the model, with varying decay exponents depending on the relative importance of rewiring and driving. Thus, this approach offers a promising avenue for investigating solar flare dynamics, particularly in relation to the distinct behaviors observed during solar minima and maxima.
	\begin{acknowledgments}
		This research was funded by FONDECyT grant No.
		1242013 (V.M. and D.P.), and ANID PhD grant No. 21231335 (A.Z.). We would also like to thank the ACAL scholarship that allowed for an internship, during which part of this work was done (A.Z.).
	\end{acknowledgments}
	
	\bibliography{sample7}{}
	\bibliographystyle{aasjournalv7}

\end{document}